\documentclass[aps,prb,amsmath,amssymb,floatfix,12pt]{revtex4-1}
\usepackage{tabularx}
\usepackage{bm}
\usepackage{euscript}
\usepackage{graphicx}
\usepackage{color}
\usepackage{amsfonts}
\usepackage{exscale}
\usepackage{amsbsy}
\usepackage{subfigure}
\usepackage{textcomp}
\usepackage{hyperref}

\usepackage{amsmath}
\usepackage{amssymb}
\usepackage{graphicx}
\usepackage{amsfonts}
\usepackage{latexsym}
\usepackage{slashed}
\usepackage{yfonts}
\usepackage{verbatim}
\usepackage{wrapfig}
\usepackage{longtable}

\numberwithin{equation}{section}
\global\newcount\itemno \global\itemno=0

\def\itemaut#1{\global\advance\itemno by1\noindent\item{\the\itemno.}#1}
\newcommand{\be}{\begin{equation}}
\newcommand{\ee}{\end{equation}}

\newcommand{\mc}{\mathcal}
\def\({\left(}
\def\){\right)}

\newcommand{\nn}{\nonumber}

\renewcommand{\th}{\theta}

\numberwithin{equation}{section}

\newcommand{\CO}{{\cal O}}

\def\IZ{{\mathbb{Z}}}
\def\IR{{\mathbb{R}}}
\def\IP{\mathbb{P}}

\def\CC {{\cal C}}
\def\CM {{\cal M}}

\def\CO {{\cal O}}

\def\CE {{\cal E}}

\def\vol{{\rm vol\,}}

\def\be{\bar{e}}

\def\half{\frac{1}{2}}

\def\be{\begin{equation}}
\def\ee{\end{equation}}

\def\t{\tau}
\def\ZZ{\mathbb Z}
\def\f{\phi}

\begin{document}

\title{Elliptic Genera and 3d Gravity}
\author{Nathan Benjamin$^{\psi, \bar\psi}$, Miranda C. N. Cheng$^{A^{\mu}}$, Shamit \nolinebreak[4] Kachru$^{\psi,\bar\psi}$, Gregory W. Moore$^{\phi}$, Natalie M. Paquette$^{\psi, \bar\psi}$}
\affiliation{$^{\psi}$Stanford Institute for Theoretical Physics, Stanford University, Stanford, CA 94305}
\affiliation{$^{\bar \psi}$SLAC National Accelerator Laboratory, 2575 Sand Hill Road, Menlo Park, CA 94025}
\affiliation{$^{A^{\mu}}$ Institute of Physics and Korteweg-de Vries Institute for Mathematics\\
University of Amsterdam, Amsterdam, the Netherlands\footnote{On leave from CNRS, France.} }
\affiliation{$^\phi$NHETC and Department of Physics and Astronomy, Rutgers University, Piscataway, NJ 08855}


\begin{abstract}

We describe general constraints on the elliptic genus of a 2d supersymmetric conformal field theory which has a gravity dual with large radius in Planck units.  We give examples of theories which do and do not satisfy the bounds
we derive, by describing the elliptic genera of symmetric product orbifolds of $K3$,
product manifolds,
certain simple families of Calabi-Yau hypersurfaces, and symmetric products of the ``Monster CFT."  We discuss the distinction between theories with
supergravity duals and those whose duals have strings at the scale set by the AdS curvature.
Under natural assumptions we attempt to quantify the fraction of (2,2) supersymmetric conformal theories which admit a weakly
curved gravity description, at large central charge.

\end{abstract}

\maketitle

\newpage
\tableofcontents
\newpage

\section{Introduction}

The AdS/CFT correspondence [\onlinecite{Maldacena}] provides a concrete framework for holography,
where very particular $d$ dimensional quantum field theories can capture the dynamics of
quantum gravity in $d+1$ spacetime dimensions.  A natural question from the outset has been:
``which class of quantum field theories is dual to (large radius, weakly coupled) Einstein gravity
theories?"

In a recent paper [\onlinecite{Hartman}], interesting progress was made on this issue in the particular
case of two-dimensional CFTs.  The authors of [\onlinecite{Hartman}] make the plausible assumption that
a weakly coupled gravitational theory should reproduce the most basic aspects of the phase structure known in all of the simple
examples of AdS/CFT.  In particular, as one raises the temperature, there should be a phase transition
at a critical temperature (usually taken to be $\beta^* = {1\over kT^*} = 2\pi$)
between a ``gas of particles" and a black hole geometry [\onlinecite{Witten}] -- the Hawking-Page transition
[\onlinecite{Hawking}].  By requiring that outside a small neighborhood of the critical temperature the
thermal partition function should be dominated by BTZ black holes at high $T$, or the ground state at
low $T$, one finds interesting constraints on the spectrum of any putative dual conformal field theory.

A significant consequence of this constraint is the derivation of the Bekenstein-Hawking black hole entropy
\be
S(E)  \sim 2\pi \sqrt{cE\over3} \;, \quad E = h+\bar h- {c\over12}
\ee
for $E >{ c\over12}$ and $c \gg 1$.
Notice that this is the regime where we expect the Bekenstein-Hawking formula to give a good approximation of the black hole entropy on gravitational grounds.  It is different from the regime of applicability of the usual Cardy formula based on familiar modular form arguments ($\frac{E}{c} \gg 1$).

Here, we turn our attention to 2d supersymmetric theories.
In two-dimensional theories with at least $(0,1)$ supersymmetry
and left and right-moving ${\mathbb Z}_2$ fermion number symmetries, one can define the elliptic genus [\onlinecite{SW,Pilch,Edold}].
We will focus on the special case of $(2,2)$ supersymmetry in this paper, but we expect that many of our considerations could be suitably generalized.
In the $(2,2)$ case, the elliptic genus associates to a 2d SCFT a weak Jacobi form; detailed knowledge of
the space of such forms (see e.g. [\onlinecite{Eichler}]) will allow us to make some strong statements
about CFT/gravity duality in this case.
Prominent cases of
such 2d supersymmetric theories in the AdS/CFT correspondence include those arising in D-brane constructions of supersymmetric black strings [\onlinecite{Strominger-Vafa}], where the near-horizon
geometry has a dual given by a $\sigma$-model with target $M^N/S_N$ for $M = K3$ or $T^4$.
By requiring the Bekenstein-Hawking formula for these black objects to apply in the black hole regime, we derive a constraint on the coefficients of the elliptic genus.

Intuitively, the condition that the CFT elliptic genus exhibits an enlarged regime of applicability of the Bekenstein-Hawking
entropy (which turns out to warrant a Hawking-Page transition) hints that there is indeed a weakly coupled gravity dual.  In the simplest perturbative string theory constructions of AdS, there are at least three scales of interest -- the
Planck scale $M_{\rm Planck}$, the string scale $M_{\rm string}$, and the inverse AdS radius $\frac{1}{\ell_{\text{AdS}}}$.  (There are also in general one or more Kaluza-Klein scales -- for simplicity we are imagining constructions like the Freund-Rubin construction where the KK scale coincides with the
AdS radius.) The
most conventional regime of understanding string models is when $M_{\rm Planck} \gg M_{\rm string} \gg {1\over \ell_{\text{AdS}}}$.  However, the conditions we impose are also satisfied in some theories where
there is no separation of scales between $M_{\rm string}$ and ${1\over \ell_{\text{AdS}}}$ apparent in the elliptic
genus.  We therefore also discuss further criteria on the coefficients of the elliptic genus which may
distinguish between theories with a separation of scales between supergravity and string modes,
and theories without such separation.

It is important to keep in mind that our necessary condition serves only as an indicator of whether there might be a weakly coupled gravity dual to some region in the moduli space of the superconformal field theory.  In simple examples, the moduli space will have other generic phases characterized by duals with no simple geometric description, and the large radius gravity dual would characterise only a small region of the SCFT moduli space.  However, as the elliptic genus is an invariant calculable (in principle) in this small
region, it will have the properties expected of a theory with a weakly coupled gravity description if the SCFT admits such a description anywhere in its moduli space.

This paper is organized as follows.  In \S2, we review some basic facts about Jacobi forms.
In \S3, we describe the constraint we wish to place on the Fourier coefficients of these
forms, following a similar philosophy to [\onlinecite{Hartman}].
In \S4, we check the bound on various simple constructions: $K3$ symmetric product orbifolds (which provide
some of the simplest examples of AdS$_3$/CFT$_2$ and do satisfy the bound), product manifolds,
a family of Calabi-Yau spaces going off to large dimension, and a symmetric product of the
``Monster CFT."
As some of the examples will fail, we see that the bound does have teeth -- there are simple examples
of (2,2) superconformal field theories at large central charge that violate it.
In \S5, focusing on the distinctions between the $K3$ symmetric product and the ``Monster" symmetric
product, we discuss the distinction between low-energy supergravity theories and low-energy string
theories.
In \S6, we attempt to quantify ``the fraction
of supersymmetric theories at large central charge which admit a gravity dual," using a natural
metric on a relevant (suitably projectivized) space of weak Jacobi forms.  Detailed arguments supporting some of the assertions in the main body of the paper are provided in two appendices.

\section{Modularity Properties}

We can define the following elliptic genera for any 2d SCFT with at least $(1,1)$ supersymmetry and
left and right-moving fermion quantum numbers.
Denote by $L_n, \bar L_n$ the left and right Virasoro generators, and $F, \bar F$ the left and right-moving fermion number.  The NS sector elliptic genus can be defined via:
\begin{equation}
Z_{NS,+}(\tau) = {\rm Tr}_{NS,R} \left(-1 \right)^{\bar F} q^{L_0 - c/24} \bar q^{\bar L_0 - c/24}~.
\end{equation}
It is a (weakly holomorphic) modular form under the congruence subgroup $\Gamma_{\theta}$, defined in (\ref{eq:gammatheta}).
Similar definitions apply in other sectors:
\begin{eqnarray}
\chi &=& {\rm Tr}_{R,R}  \left(-1\right)^{F + \bar F} q^{L_0 - c/24} \bar q^{\bar L_0 - c/24}\\
Z_{R,+}(\tau) &=& {\rm Tr}_{R,R} \left( -1\right)^{\bar F} q^{L_0 - c/24} \bar q^{\bar L_0 - c/24}\\
Z_{NS,-}(\tau) &=& {\rm Tr}_{NS,R}\left(-1\right)^{F + \bar F} q^{L_0 - c/24} \bar q^{\bar L_0 - c/24}~.
\end{eqnarray}
Here, $q = e^{2\pi i \tau}$ where $\tau$ takes values in the upper half plane, and we have assumed equal left and right-moving central charges,
$c_L = c_R = c$.

For the most part, we will consider theories with additional structure, e.g. (2,2) superconformal
theories.  In fact for any (0,2) theory with a left-moving $U(1)$ symmetry, and so in particular for
any (2,2) SCFT, one can define a refined
elliptic genus as
\be
Z_{R,R}(\t,z) =  {\rm Tr}_{R,R} (-1)^{F+\bar F}q^{L_0 - c/24} \bar q^{\bar L_0 - c/24}y^{J_0}.
\label{partfun}
\ee
Here, $y = e^{2\pi i z}$.  The additional symmetry promotes the two-variable elliptic genus into a weak
Jacobi form [\onlinecite{Yang}].
We will also consider
\begin{align}
Z_{NS,R}(\t,z) &= {\rm Tr}_{NS,R} (-1)^{\bar F} q^{L_0 - c/24} \bar q^{\bar L_0 - c/24} y^{J_0} \nn \\
&= Z_{R,R}\(\tau, z + \frac{\tau+1}{2}\) q^{\frac{c}{24}} y^{\frac{c}{6}}.
\end{align}
Note that we could define $Z_{NS,NS}$ as a quantity which localizes on right-moving chiral primaries, but
with suitable definition it would give the same function as $Z_{NS,R}$ above.  So, while the
AdS vacuum appears in the $(\text{NS},\text{NS})$ sector,  we will
focus on $Z_{NS,R}$ when stating our bounds in \S3.

In the cases of interest to us, there is no anti-holomorphic dependence on $\bar\tau$ due to the   $(-1)^{\bar F}$ insertion, and the elliptic genus is a purely holomorphic function of
$\tau$.  In fact, much more is true.  Using standard arguments one can show that the elliptic genus of a SCFT defined above in (\ref{partfun}) transforms nicely under the group $\ZZ^2 \rtimes SL(2,\ZZ)$. In particular, it is a so-called weak Jacobi form of weight 0 and index $c/6$, defined below. For instance, supersymmetric sigma models for Calabi-Yau target spaces of complex dimension $2m$ have elliptic genera that are weight 0 weak Jacobi form of index $m$. For the rest of this paper, we will be considering SCFTs with $m \in \mathbb{Z}$, or equivalently $c$ divisible by $6$.

Consider a holomorphic function $\phi(\tau,z)$ on
${\mathbb H} \times {\mathbb C}$ which satisfies the conditions
\begin{align}\label{modular_transformation}
\phi\({a\tau + b \over {c\tau + d}}, {z\over c\tau + d}\) &= (c\tau + d)^w e^{2\pi i m {cz^2 \over c\tau + d}} \phi(\tau,z)\;, \;\begin{pmatrix} a&b \\ c&d \end{pmatrix} \in SL_2({\mathbb Z}) \\
\label{elliptic_transformation}
\phi(\tau, z + \ell \tau + \ell^\prime) &= e^{-2\pi i m (\ell^2\tau + 2\ell z)}\phi(\tau,z)\;, \;
\ell,\ell^\prime \in {\mathbb Z}~.
\end{align}
In the present context, (\ref{elliptic_transformation}) can be understood in terms of the spectral flow symmetry in the presence of an ${\cal N}\geq2$ superconformal symmetry.

The invariance  $\f(\tau,z)=\f(\tau+1,z)=\f(\tau,z+1)$ implies a Fourier expansion
\be\label{eqn:forms:jac:FouExp}
\phi(\t,z) = \sum_{n,\ell \in \ZZ} c(n,\ell) q^n y^\ell ,
\ee
and the transformation under $(\begin{smallmatrix} -1 & 0 \\ 0 &-1\end{smallmatrix})\in SL_2(\ZZ)$ shows
\begin{equation}
c(n,\ell) = (-1)^w c(n,-\ell).
\end{equation}
The function $\phi(\tau,z)$ is called a weak Jacobi form of index $m \in {\mathbb Z}$ and weight $w$ if its Fourier coefficients $c(n, \ell)$ vanish for $n<0$.
Moreover, the elliptic transformation (\ref{elliptic_transformation}) can be used to show that
the coefficients
\be \label{jac_coeff}c(n,\ell) = C_{r}(D(n,\ell))\ee
depend only on the so-called discriminant
\be
D(n,\ell):= \ell^2-4mn
\label{eq:d}
\ee
and $r= \ell \pmod{2m}$. Note that $D(n,\ell)$ is the negative of the polarity, defined in [\onlinecite{Gaberdiel}] as $4mn-\ell^2$.

Combining the above, we see that a Jacobi form admits the expansion
\be\label{eqn:forms:jac:thetaxpn}
\phi(\tau,z)= \sum_{r \in \ZZ/2m\ZZ}   h_{m,r}(\tau) \th_{m,r}(\tau,z)
\ee
in terms of the index $m$ theta functions,
\be\label{thetexp}
\theta_{m,r}(\tau,z) =\sum_{\substack{k= r~{\text{mod}}~2m}} q^{k^2/4m} y^{k}.
\ee
Both $h_{m,r}$ and $\theta_{m,r}$ only depend on the value of $r$ modulo $2m$. However,
for some later manipulations we should note that it is sometimes useful to choose the
explicit fundamental domain  $-m~<~r\leq~m$ for the shift symmetry in $r$.
When $\vert r \vert \leq m$  we can write:
\be
h_{m,r}(\tau)  = (-1)^w h_{m,-r}(\tau) =\sum_{n\geq0} c(n,r)   q^{-D(n,r)/4m} .
\ee
The vector-valued functions $\th_{m, r}(\t, z)$ transform as
\begin{align}
\theta_{m}(-\frac{1}{\t},-\frac{z}{\t}) &= \sqrt{-i \t}\, e^{\frac{2\pi i m z^2}{\tau}}\, {\cal S} \,\theta_{m}(\t,z),\qquad\label{transf_theta}\\
\theta_{m}({\t}+1,{z}) &= {\cal T} \,\theta_{m}(\t,z),
\end{align}
where ${\cal S}$, ${\cal T}$ are the $2m \times 2m$ unitary matrices with entries
\begin{align}
{\cal S}_{rr'} &= \frac{1}{\sqrt{2m}} e^{\frac{\pi i r r'}{m}},\\ \label{transf_theta_2}
{\cal T}_{rr'} &= e^{\frac{\pi i r^2}{2m}} \delta_{r,r'}.
\end{align}
From this we see that $ h= (  h_{m,r})$ is a $2m$-component vector transforming as a weight $w-1/2$ modular form for $SL_2(\ZZ)$.

In particular, an elliptic genus  (with $w=0$) of a theory with central charge $c=6m$ can be written as
\be\label{rth_EG}
Z_{R,R}(\tau,z) = \sum_{r\in \ZZ/2m\ZZ} Z_r(\tau) \theta_{m,r}(\tau,z).
\ee
We have written $Z_r(\tau)$ for $h_{m,r}(\tau)$ in this expression. Thus $Z_r(\tau)$ only depends on $r$ modulo $2m$,
but again,  when $\vert r \vert \leq m$ it is useful to
expand:
\be
Z_r(\tau) = Z_{-r}(\tau) =\sum_{n\geq 0} c(n,r) q^{n-\frac{r^2}{4m}}
\ee
The function $Z_r(\tau)$ can be thought of as the elliptic genus of the $r^{\text{th}}$ superselection sector corresponding to the eigenvalue of $J_0=r \mod{2m}$. From the CFT point of view, the $r\sim r + 2m$ identification can be understood in terms of the spectral flow symmetry of the superconformal algebra. When there is a gravity dual the $r\to r+2m$ transformation corresponds, from the
bulk viewpoint,
to a large gauge transformation of a gauge field holographically dual to the $U(1)_R$.

Since the Fourier coefficients of a weak Jacobi form have to satisfy
\be
c(n,\ell)=0 \quad{\text{ for all }}\quad n<0,
\ee
(which can be thought of as unitarity of the CFT), this leaves open the possibility to have ``polar terms" $c(n,\ell)q^n y^\ell$ with
\[
m^2 \geq D(n,\ell) > 0~,~n\geq0
\]
in an index $m$ weak Jacobi form. These are called polar terms because they are precisely the terms in the $q$-series of $Z_{r}(\tau)$ that have exponential growth when approaching the cusp $\tau \to i\infty$. The finite set of independent coefficients of the polar terms in the
elliptic genus will play a crucial role in what follows. In what follows we will denote by $\phi_P$ the sum of all the polar terms in the
elliptic genus.

Importantly, the full set of Fourier coefficients of a weak Jacobi form can be reconstructed from
just the polar part, $\phi_P$.
This can be understood through the fact that there are no non-vanishing negative weight modular forms at any level.
For discussions of this in related contexts, see [\onlinecite{Gaberdiel, FaeryTail, ModernFaeryTail}].
Let us denote by $V_m$ the space of possible polar polynomials (without requiring that they
correspond to the polar part of a bona fide weak Jacobi form).
Given the symmetries of the $c(n,\ell)$, $V_m$ is spanned by $q^n y^\ell$ in the region
${\cal P}^{(m)}$:
\begin{equation}
{\cal P}^{(m)} = \{ (\ell,n): 1 \leq \ell \leq m,~~0 \leq n,~~D(n,\ell) > 0 \}~.
\end{equation}
By a standard counting of the number of lattice points underneath the parabola $4 mn - \ell^2 =0$
in the $\ell, n$ plane [\onlinecite{Gaberdiel}], one can give a formula for
the dimension of the vector space of polar parts $P(m) = {\rm dim}(V_m)$:
\begin{equation}
\label{dimis}
P(m) = \sum_{\ell = 1}^m \lceil {\ell^2 \over 4m} \rceil~.
\end{equation}
In this note, where we work at leading order in large $m$, we will only need the leading behavior of
the sum (\ref{dimis}); this is determined by the elementary formula
$\sum_{\ell=1}^m {\ell^2} = {1\over 3}m^3 + {1\over 2}m^2 + {1\over 6} m$ to be
\begin{equation}
\label{pig}
P(m) = {1\over 12} m^2 + O(m),~~m \gg 1~.
\end{equation}

Because we are working at leading order at large $m$ (large central charge), we will not need to
use the subleading corrections to (\ref{pig}) (determined in [\onlinecite{Gaberdiel}]).  Neither will we need
need to deal with the important subtlety that not all vectors in $V_m$ actually correspond to a
weak Jacobi form.  Denoting the space of weak Jacobi forms of weight $0$ and index $m$ as
$\tilde J_{0,m}$, in fact one has ${\rm dim}(\tilde J_{0,m}) - P(m) = O(m)$. These facts would become important if one were to extend our results to the
next order in a $1/m$ expansion.

\section{Gravity constraints and Phase structure}

We will now derive a constraint on the polar coefficients of a SCFT as follows.  The polar coefficients determine the elliptic genus, and we will require that the genus
matches the expected Bekenstein-Hawking entropy of black holes in the high-energy regime.
Happily, we will find that a second (a priori independent) requirement of the existence of a sharp Hawking-Page transition at
the critical temperature $\beta = 2\pi$ gives the same constraint on the coefficients.

More precisely, we will be
considering infinite sequences of CFTs going off to large central charge, and we will bound the
asymptotic behavior of physical observables in such sequences as $m \to \infty$.  (One familiar example that can be taken as representative of what we have in mind is the sequence of $\sigma$-models
with targets ${\rm Sym}^{m}(K3)$.)
Simple physical considerations will lead us to propose certain constraints on the growth of the polar coefficients at large $m$ in the related families of elliptic genera.

Now, there are precise mathematical statements on the behaviors of coefficients of large powers of $q$ in modular forms.
For instance, there are theorems proving that for a generic holomorphic modular form
$f = \sum_n c_n q^n$ of fixed weight $k$, $c_n$ grows as $O(n^{k-1})$ at large $n$, while
for a cusp form, the coefficients are of $O(n^{k/2})$.

Note that our growth estimates are rather different in nature from those of the previous paragraph.
Our estimates will be $\it physically$ motivated by known facts about
corrections to Einstein gravity in the expansion in energy divided by $M_{\rm Planck}$.
We are proposing a
mathematical criterion, motivated by physics, that would allow one to check whether a given
sequence of CFTs can possible have a weakly coupled gravity dual.
This could equally well be viewed as a mathematical conjecture about the families of modular
forms arising in sequences with gravity duals.

Our eventual criterion will be derived by considering the free energies $F_m$ of the CFTs in this family.
 The free energy in these theories, as
$m \to \infty$, gives a function with a sharp first-order phase transition at $\beta = 2\pi$. This is the physical phenomenon of the Hawking-Page transition [\onlinecite{Hawking}].
(Sharp roughly because, in microscopic examples of AdS$_3$/CFT$_2$, semi-classical configurations of winding strings can
condense and lower the free energy precisely at $2\pi$, yielding the transition -- see [\S5.3.2, \onlinecite{review}]).  Similarly, when we state physically motivated criteria about
the free energies of our sequences of theories, we will be making statements about the sequence
$F_m$ and assuming that the limit as $m \to \infty$ of ${1\over m} F_m$ exists as a piece-wise differentiable function
with discontinuous first-derivative at $\beta=2\pi$.

\subsection{A Bekenstein-Hawking bound on the elliptic genus}

Suppose that $\phi$ is the elliptic genus
of a superconformal field theory with a large radius gravitational dual.  Define the ``reduced
mass" of a particle state in the dual gravity picture to be the eigenvalue of
\begin{equation}
L_0^{\rm red} = L_0 - {1\over 4m} J_0^2 - {m\over 4}~,
\label{eq:l0defn}
\end{equation}
namely the quantity ${-D(n,\ell)}/{4m}$ for the term $q^n y^\ell$ in the elliptic genus. Define $E^{\text{red}}$ to be the eigenvalue under $L_0^{\text{red}}$.
Then:

\noindent
$\bullet$
Classically, the states with $E^{\text{red}} > 0$ are black holes in AdS$_3$.  We will discuss their contribution
to the supergravity computation of the elliptic genus in detail below.

\noindent
$\bullet$
In contrast, in the gravitational computation of the elliptic genus, it is the states with
$E^{\text{red}} < 0$ which contribute to the polar part of the supergravity partition function
[\onlinecite{FaeryTail}].  These are precisely the modes which are too light to form black
holes in the bulk.  These  are the states which appear in $\phi_P$.

We now present an argument that constrains the coefficients in $\phi_P$ using the
supergravity estimate of the black hole contribution to the elliptic genus.
We treat the elliptic genus as the grand canonical partition function
\be
Z(\beta,\mu)= \sum_{\rm{microstates}} e^{\beta(\mu Q -E)} =e^{-\beta F(\beta,\mu) },
\ee
where $\tau=i {\beta\over 2\pi }$ and $z=- i {\beta \mu\over 2\pi }$ are the corresponding variables in the elliptic genus.
In other words, we define
\be\label{identification}
Z(\beta,\mu)=Z_{NS,R}(\tau=i {\beta\over 2\pi }, z=- i {\beta \mu\over 2\pi }).
\ee
To make contact with the usual thermodynamical analysis, we will require $\beta$ and $\mu$ to be real numbers.
Let us discuss the supergravity estimate for this in simple steps.  See also, for instance, the nice discussions in [\onlinecite{Kraus:2006nb}, \onlinecite{Kraus:2006wn}].

\subsubsection{Uncharged BTZ}

In calculating the elliptic genus for a 2d SCFT, we restrict to states that are ground states on the right-moving side, but with arbitrary $L_0$. These correspond to extremal spinning black holes in the 3d bulk, with vanishing
temperature $T=0$.

We can calculate the entropy of these black holes using the standard properties of black hole thermodynamics [\onlinecite{BTZ}]. We will work in units where $\ell_{\text{AdS}}=1$. The inner and outer horizons coincide for the extremal geometries, and are located at
\begin{align}
\label{superpig}
r_+ = r_- &= 2 \sqrt{G M}~.
 \end{align}
The entropy is given by
\begin{align}
S&=\frac{\pi r_+}{2G} ~.
\end{align}
Finally, the central charge of the Brown-Henneaux Virasoso algebra is related to $G$ by
\begin{align}
c=\frac{3}{2G}.
\end{align}
Combining, we get
\begin{align}
S = 2\pi\sqrt{\frac{c M}6}.
\end{align}
If we were to include Planck-suppressed corrections to the black hole entropy,
we expect no fractional powers of $M_{\rm Planck}$ to appear in the corrected formula, but
corrections which involve ${\rm log}(M_{\rm Planck})$ can appear.  This translates into
$O(\log{c})$ corrections, but no power-law in $c$ corrections, to the entropy.

The black hole mass $M$ is identified with the eigenvalue of $L_0 - {c\over 24}$, which we will denote as $n$. This means that the degeneracy of states of the elliptic genus $c_n$ goes as
\begin{align}
c_n = e^{2\pi \sqrt{\frac{c n}6} + O(\log{n})}~.
\end{align}
This is the familiar Cardy-like growth. As we are interested in studying families of CFTs asymptoting
to the large central charge limit, we would like to know about the behavior at fixed $n$ as $c \to \infty$.  For this purpose, the
more informative expression would be
\begin{equation}
\label{also}
c_n = e^{2\pi \sqrt{cn \over 6} + O(\log{c})}~.
\end{equation}

As an aside, let us discuss the validity of the above equation.  The above derivation of the black hole contribution to
the partition function is valid whenever the radius of the black hole is large in Planck units.
The first BTZ black hole appears at a mass $\sim M_{\rm Planck}$, and we see from
(\ref{superpig}) that its radius will already be quite large -- of order $\ell_{AdS}$, or $O(c)$ in
Planck units.  We then expect the semi-classical entropy formula to be valid for even very light
black holes at large $c$. This is one way to understand the characteristic Cardy-like growth of the number of states of CFTs with gravity duals, even outside the usual range of validity of the Cardy formula that is guaranteed by modular invariance alone.

Writing the elliptic genus now as
\begin{align}
Z(\tau) = \int dn ~e^{2\pi \sqrt{\frac{cn}6}} e^{2\pi i \tau n}
\end{align}
we can ask the question: at fixed $\tau$ (where $\frac{i}{2\pi\tau} = \frac{1}{\beta}$ is the formal ``temperature" variable; not to be confused with the temperature of the black hole, which is zero), what value of $n$ dominates the sum? This is solved using standard saddle point approximation methods. The derivative of $e^{2\pi \sqrt{\frac{cn}6}} e^{2\pi i \tau n}$ vanishes when
\begin{align}
2\pi i \tau = -\pi \sqrt{\frac{c}{6n}}
\end{align}
or equivalently
\begin{align}
\beta = \pi \sqrt{\frac{c}{6n}}.
\end{align}
Thus we get
\begin{align}
n = \pi^2  {m\over \beta^2}.
\end{align}
Using the famous relation $F = E - TS$, and recalling that the temperature of the black hole is 0, we therefore get
\begin{align}
F = -\pi^2 {m\over \beta^2} + O(\log{m}).
\end{align}
We were careful to write $\beta$ here to distinguish from the physical $T=0$ of the extremal black holes contributing to the genus.  (While the torus partition function at a given $\tau$ would correspond to a thermal ensemble, the elliptic genus is only counting extremal states and the temperature represented by ${\rm Im}(\tau)$ is fictitious.)

\subsubsection{Adding Wilson lines}

Now we turn to the elliptic genus, a refinement of the above discussion which keeps track of $U(1)$ charge.

In the bulk, the existence of the $U(1)_R$ symmetry of the dual (2,2) SCFT is manifested in the presence of Chern-Simons gauge fields.  First, let us discuss the expected effect heuristically.
By adding a $U(1)$ Chern-Simons gauge interaction at level $k$, we add to the action the following boundary term
\be
S_{gauge}^{bdry}=-\frac{k}{16\pi} \int_{\partial AdS} d^2x \sqrt{g} g^{\alpha\beta} A_{\alpha} A_{\beta}.
\ee
For a BTZ black hole, the angular direction in the 2d spatial manifold (which we shall call the $\phi$ direction) is non-contractible, so we allow $A_{\phi}$ to be nonzero.

We thus shift the action by a term proportional to $A^2$.  This will add a term that goes as $\mu^2$ to the free energy so we will get something like
\be
F \sim {m \over \beta^2} + k \mu^2~.
\ee
Finally, for a (2,2) SCFT with $k$ determined by the central charge and hence the index $m$,
we will have
\begin{equation}
F\sim {m \over \beta^2} + m \mu^2~.
\end{equation}

Now, let's be more explicit.  The entropy of the black holes we are considering is given, in general, by
[\onlinecite{Cvetic}]
\begin{align}
S &= 2\pi \sqrt{m} \sqrt{n - \frac{\ell^2}{4m}} \nn \\
&= \pi \sqrt{-D(n,\ell)}
\label{BHentropy}
\end{align}
where $n$ is the eigenvalue under $L_0-{c\over 24}$, and $\ell$ is the $J_0$ eigenvalue.

Now, again, we write the degeneracy
\be
c(n, \ell) = e^{2\pi \sqrt{m} \sqrt{n - \frac{\ell^2}{4m}} + O(\log{(n-\frac{\ell^2}{4m}}))},
\ee
or following the analogous discussion above
\be
c(n, \ell) = e^{2\pi \sqrt{m} \sqrt{n - \frac{\ell^2}{4m}} + O(\log{m})},
\ee
and the elliptic genus can be written as
\be
Z_{NS,R}(\tau, z) = \int dn \int d\ell ~e^{2\pi i \tau n} e^{2\pi i z \ell} e^{2\pi \sqrt{m} \sqrt{n - \frac{\ell^2}{4m}}}~.
\ee
This has a saddle when
\begin{align}
\tau &= \frac{im}{\sqrt{4mn-\ell^2}} \nn\\
z &= \frac{-i\ell}{2\sqrt{4mn-\ell^2}}.
\end{align}
Rewriting, the dominant saddle occurs at
\begin{align}
n&=m({\pi^2 \over \beta^2} + \mu^2)\nn\\
\ell&=2m\mu.
\end{align}
Thus, we get the free energy as
\be
\label{omigod}
F = -m{\pi^2 \over \beta^2} - m \mu^2 + O(\log{m}).
\ee

Identifying this free energy with $-{1\over \beta} {\rm log} Z$ gives us the behavior
of the elliptic genus.
However, we need to be sure that the supergravity derivation is valid -- i.e.
that the configurations we included correspond to reliable and dominant saddle points.
Reliability follows if the black hole is large in Planck units, which works for any $E^{\text{red}} > 0$ at large $c$.  We also require that the black hole saddle be the dominant one.
This will be true for any $\beta < 2\pi$ at very large $m$.  For $\beta > 2\pi$, instead the
``gas of gravitons" dominates, and (\ref{omigod}) is not the appropriate expression for the
free energy.  Finally, in a tiny neighborhood of $\beta = 2\pi$, the free energy crosses from the
value for the gas of gravitons to the value characteristic of black holes above; this is a regime
where ``enigma black holes" play an important role, and cannot be characterized in a universal way.
In known microscopic examples of AdS$_3$/CFT$_2$, these are small black holes  (localized on the transverse sphere) of negative specific heat
(see e.g. [\onlinecite{deBoer:2008fk}, \onlinecite{Bena:2011zw}] for discussions).

Next we will derive constraints on the low-temperature expansion -- and in particular the polar
coefficients -- from these results of black hole thermodynamics.

\subsubsection{Bounds on polar coefficients}

After these physical preliminaries we are ready to derive the main result of this paper.
This result will follow (given appropriate physical assumptions) by
combining  modular invariance with the physical requirement that $Z(\beta,\mu)$
has large $m$ asymptotics given by
\be
\log Z(\beta,\mu) =  m\bigl( {\pi^2 \over \beta} +   \beta \mu^2 \bigr) + O(\log{m}),
\label{eq:reproducethis}
\ee
for all real $(\beta, \mu)$ such that $0< \beta < 2\pi$. Recall from
equation \eqref{identification} that $Z(\beta,\mu)$ is just the elliptic genus $Z_{NS,R}(\tau,z)$
evaluated for  $\tau =i\beta/2\pi $ and $z=-i\beta\mu/2\pi$.

Now we write out the modular property:
\begin{align}
Z_{NS,R}(\tau,z) &= (-1)^m e^{-\frac{2\pi i mz^2}{\tau}}  Z_{NS,R}(-\frac{1}{\tau},-\frac{z}{\tau}).
\label{eq:modularitybonanza}
\end{align}
We make a few elementary manipulations:
\begin{align}
Z_{NS,R}(\tau, z) &= e^{\frac{2\pi i \tau m}{4}} e^{2\pi i z m} Z_{R,R}(\tau, z + \frac{\tau+1}{2}) \nn \\
&= e^{\frac{\pi i \tau m}{2}} e^{2\pi i z m} \sum_{\substack{r \in \mathbb{Z}/2m\mathbb{Z}}} Z_r(\tau) \theta_{m,r}(\tau, z + \frac{\tau+1}{2}) \nn \\
&= e^{\frac{\pi i \tau m}{2}} e^{2\pi i z m} \sum_{\substack{r \in \mathbb{Z}/2m\mathbb{Z}}} \sum_{\substack{D\leq r^2 \\ D = r^2 \text{~mod~}4m}} C_r(D) e^{-\frac{2\pi i \tau D}{4m}} \sum_{k = r \text{~mod~}2m} e^{\frac{2\pi i k^2 \tau}{4m}} e^{2\pi i z k} (-1)^k e^{i \pi \tau k} \nn \\
&= e^{-\frac{\pi i \tau m}{2}} e^{2\pi i z m} \sum_{\substack{r \in \mathbb{Z}/2m\mathbb{Z}}} \sum_{\substack{D\leq r^2 \\ D = r^2 \text{~mod~}4m}}  \sum_{k = r \text{~mod~} 2m} C_r(D) e^{\frac{2\pi i \tau}{4m}(-D + m^2 + (k+m)^2)} e^{2\pi i z k} (-1)^k.
\label{eq:bonanzapartdeux}
\end{align}
Combining (\ref{eq:modularitybonanza}) and (\ref{eq:bonanzapartdeux}), we get
\begin{align}
Z_{NS,R}(\tau,z) &=  e^{-\frac{2\pi i mz^2}{\tau}} e^{\frac{i \pi m}{2\tau}} e^{-\frac{2\pi i z m}{\tau}} \nn \\&\phantom{aaaaaaa}\sum_{\substack{r \in \mathbb{Z}/2m\mathbb{Z}}} \sum_{\substack{D\leq r^2 \\ D = r^2 \text{~mod~}4m}}  \sum_{k = r \text{~mod~}2m} C_r(D) e^{\frac{2\pi i}{4m\tau}(D - m^2 - (k+m)^2)} e^{-\frac{2\pi i z k}{\tau}} (-1)^{k+m} \nn \\
&= e^{m \beta \mu^2 + \frac{m \pi^2}{\beta}} \sum_{\substack{r \in \mathbb{Z}/2m\mathbb{Z}}} \sum_{\substack{D\leq r^2 \\ D = r^2 \text{~mod~}4m}}  \sum_{k = r \text{~mod~}2m} C_r(D) e^{\frac{\pi^2}{m\beta}(D-m^2-(k+m)^2)} e^{2\pi i (k+m) (\mu+\frac{1}{2})}
\label{eq:barf}
\end{align}
where in the last line we have used the substitutions $\tau =\frac{i\beta}{2\pi}$ and $z=-\frac{i\beta\mu}{2\pi}$.

Note that the prefactor in front of the sum in equation \eqref{eq:barf} gives the right hand
side of equation \eqref{eq:reproducethis}.  Therefore

 \begin{align}
{\rm log} \left( \sum_{\substack{r \in \mathbb{Z}/2m\mathbb{Z}}} \sum_{\substack{D\leq r^2 \\ D = r^2 \text{~mod~}4m}}  \sum_{k = r \text{~mod~}2m} C_r(D) e^{\frac{\pi^2}{m\beta}(D-m^2-(k+m)^2)} e^{2\pi i (k+m) (\mu+\frac{1}{2}) } \right) &\sim O(\log(m)). \nn \\
 \end{align}

In order to turn this into a more useful statement we next
 introduce another physically motivated hypothesis -- the ``non-cancellation hypothesis.''
This hypothesis states that the leading order large $m$
asymptotics is not affected if we replace the terms in the expansion of $Z_{NS,R}$ above
 by their \emph{absolute values}\footnote{Since $Z_{NS,R}$ is modular this again can only be valid in a distinguished set of
expansions around cusps,  and we take it to apply to the expansion in (\ref{eq:barf}).}.
Given the noncancellation hypothesis none of the terms in the sum can get large, and hence we arrive at
the necessary condition:
\be
\text{~~~~~~}\log{\(\vert C_r(D)\vert e^{\frac{\pi^2}{m\beta}(D - m^2 - (k+m)^2)}\)} =  O(\log{m}) \text{~~~~for all~}\beta<2\pi \text{~and~}k=r\text{~mod~}2m.
\ee

The strongest bound is obtained by taking the limit as $\beta$ increases to $2\pi$ from below,
yielding:
\be
|C_r(D)| \leq e^{\frac{2\pi}{4m}(m^2 - D + \min{\lbrace(k+m)^2 \vert k = r (\text{mod~}2m)\rbrace}) + O(\log{m})}.
\ee
We can write the bound simply in terms of coefficients $c(n,\ell)$ where $0\leq \ell \leq m$; the rest of the coefficients will be determined from this subset via spectral flow and reflection of $\ell$. We then get the bound
\be
\boxed{|c(n,\ell)| \leq e^{2\pi\(n + \frac{m}{2}-\frac{|\ell|}{2}\)+O(\log{m})}.}
\label{thebound}
\ee
Put differently, $|e^{-2\pi\(n + \frac{m}{2}-\frac{|\ell|}{2}\)} c(n,\ell)|$ can grow at most as a
power of $m$ for $m\to \infty$. In addition to these conditions, the bound should not be saturated
by an exponentially large number of states.

We conclude with a few remarks.

\begin{enumerate}

\item To be fastidious, the bound \eqref{thebound} applies to any family $\CC^{(m)}$ of CFTs with a
weakly coupled gravity dual, together with a sequence $(n(m), \ell(m))$ of lattice points
such that the sequence of elliptic genus coefficients
$c(n(m), \ell(m); \CC^{(m)}  )$ has well-defined large $m$ asymptotics.

\item The $O(\log{m})$ error term in the exponent can be understood in various ways.  Perhaps the most enlightening physically is that it can be directly connected (via modularity) to the  $M_{\rm Planck}$ suppressed corrections to the black hole entropy in the
$\beta < 2\pi$ regime.

\item Note that the bound is already nontrivial for the coefficient $c(0,m)$ of the extreme polar term
with $(n,\ell)=(0,\pm m)$. Under spectral flow the states contributing to this degeneracy
correspond to the unique NS-sector vacuum on the left tensored with one of the
 Ramond sector ground states on the right. We will see that already the bound on the extreme polar states is
useful.

\item  Notice that in (\ref{BHentropy}), we have only written a formula for the entropy in the stable black
hole region $E^{\rm red} > {c\over 24}$.  This follows because our saddle point approximation is only self-consistent when $\beta < 2\pi$ in this range of energies. While it may seem naively that the large $c$ behavior of
the free energy would guarantee this formula also for $0 < E^{\rm red} < {c\over 24}$, this is not
the case.  Because there is a jump of $O(c)$ in the energy in a small neighborhood of $\beta =
2\pi$, in this window $O(1)$ contributions to the free energy (which we've neglected in the
large $c$ limit) could lead to significant changes in $E^{\text{red}}$; our formula for $S(E^{\text{red}})$ is then
unreliable.  It becomes reliable once one reaches the stable range of energies $E^{\text{red}} > {c\over 24}$.
For further discussion of this issue, see [\onlinecite{Hartman}] as well as  [\onlinecite{deBoer:2008fk}, \onlinecite{Bena:2011zw}].

\end{enumerate}

\subsection{On the Hawking-Page transition}

In what follows we will present an alternate derivation of (\ref{thebound}) by insisting on a sharp Hawking-Page phase transition near $\beta = 2\pi$ (in the limit of large central charge) in the NS-R sector.  The sharp transition is not a surprise.  It is expected from general properties of the AdS$_3$/CFT$_2$ duality (and in particular, from the existence of light multiply-wound strings which can lower the free energy once $\beta < 2\pi$, in known microscopic examples [\S5.3.2, \onlinecite{review}]).

Recall that the NS sector elliptic genus has a $q$-expansion of the form
\begin{equation}
\label{hog}
Z_{NS,+}(\tau) =   q^{m\over 4} Z_{RR}\big(\t,\frac{\t+1}{2}\big) =  \sum_{n,\ell}(-1)^{\ell} c(n,\ell) \, q^{{m\over 4} + n + {\ell \over 2}}~.
\end{equation}
From the modular properties of $Z_{RR}(\t,z)$ we see that  $Z_{NS,+}(\tau) $ is invariant under the group
\be
\Gamma_\theta = \left\{ \begin{pmatrix} a & b\\ c&d\end{pmatrix}\in SL_2(\ZZ) \Big\lvert\, c-d \equiv a-b \equiv 1\pmod{2} \right\}
\label{eq:gammatheta}
\ee
which is conjugate to the Hecke congruence group $\Gamma_0(2)$.

Clearly, it satisfies at the lowest temperatures
\begin{equation}
{\rm log} Z_{NS,+}(\tau = i {\beta \over 2\pi}) = {c\over 24} \beta,~~\beta \gg 2\pi ~.
\end{equation}
To have a phase dominated by the ground state until temperatures parametrically close to
$\beta =  2\pi$ at large central charge $c = 6m$, one requires:
\begin{equation}
\label{GSdominates}
{\rm log} Z_{NS,+}(\tau= i {\beta \over 2\pi}) = {c\over 24}\beta + O(\log c),~\beta > 2\pi.
\end{equation}
Again, this can be viewed as an asymptotic condition on a family of CFTs which has a weakly
curved gravity dual
at large $m$:
the limit as $m \to \infty$ of ${1\over m} \log{Z_{NS,+}}$ for any $\beta > 2\pi$ exists and
asymptotes to ${1\over 4} \beta$.

The size of the sub-leading terms in (\ref{GSdominates}) requires some discussion.  In fact, just for the purpose of  having a phase transition at $\beta=2\pi$ in the large $c$ limit, it is possible to relax the condition of strict ground state dominance and to allow ${\rm log} Z_{NS,+}(\beta) = {c \over 24}\beta + O(c^{1-\delta})$ for some $\delta > 0$, instead of restricting to $O(\log c)$.
As noted before, however, in the large temperature regime this would imply corrections to the Bekenstein-Hawking entropy suppressed by fractional powers of $M_{\rm Planck}$, which are not expected. On the other hand logarithmic corrections are expected. This suggests one should set $\delta = 1$.
In any case, we shall not pursue the slight generalization to $\delta \neq 0$ in the present paper -- the requisite modification of the analysis can be implemented in a relatively straightforward way.

A sufficient condition for (\ref{GSdominates}) to be true is that
$\vert c(n,\ell) q^{\frac{m}{4} + n + \frac{\ell}{2} }\vert \leq e^{m\beta/4} $
for a number of terms which grows at most polynomially in $m$. If we invoke
the noncancellation hypothesis we can also say that a necessary condition is:
 \begin{equation}
\label{deltaonebound}
\vert c(n,\ell)\vert \leq  e^{2\pi (n - \frac{|\ell|}{2} + \frac{m}{2}) +  O(\log m)},
\end{equation}
If we combine this statement with the spectral flow property
$c(n,\ell) = c(n+s\ell+ms^2, \ell +2 sm )$ for all integers $s$ we can
get the best bound by minimizing with respect to $s$, subject to the condition
that $s$ is integer. Combining with reflection invariance on $\ell$ it
is not difficult to show then that the best bound is
 \begin{equation}
\label{deltaoneboundv2}
\vert c(n,\ell)\vert \leq  e^{2\pi (n_0 - \frac{|\ell_0|}{2} + \frac{m}{2}) +  O(\log m)},
\end{equation}
where $(n,\ell)$ is related to $(n_0,\ell_0)$ by spectral flow and reflection and
$0\leq \ell_0\leq m$. This is
the same condition we have derived to reproduce Bekenstein-Hawking entropy (\ref{thebound}).

The above phase transition corresponds to moving between ${\rm Im}(\tau) = 1-\epsilon$ and ${\rm Im}(\tau) = 1+ \epsilon$ with ${\rm Re}(\tau)=0$ between two specific copies of the fundamental domain of $\Gamma_\theta$. See Figure 1.
In the Euclidean signature, other saddle points corresponding to analytic continuation of the BTZ black holes are also believed to be relevant [\onlinecite{FaeryTail}, \onlinecite{exclusion}], and one is led to a stronger prediction for a phase diagram requiring an infinite number of different phases corresponding to pairs $(c, d)$ of co-prime integers with $c\geq0$, $c-d\equiv 1\pmod{2}$ (see [\onlinecite{FaeryTail}] and [\onlinecite{Maloney:2007ud}, \S 7.3])\footnote{Reference [\onlinecite{FaeryTail}] erroneously claimed the phase diagram would be
invariant under $PSL(2,\IZ)$. However the argument given there is easily corrected, and it
predicts a phase diagram invariant under $\Gamma_0(2)$ for the NS-sector genus considered there.
For further discussion see Appendix A.}.

One should then obtain a phase structure which divides the upper half-plane into regions dominated
by the various saddle points labelled by different values of $(c,d)$.
This corresponds to a tessellation of the upper-half plane by $\Gamma_\infty\backslash \Gamma_\theta$ where $\Gamma_\infty$ is the group generated by $T^2$, coinciding with the intersection of $\Gamma_\theta$ and $\langle T \rangle$. This tessellation is drawn in Figure 1 with the thick lines.
We discuss the derivation of this phase diagram in detail in Appendix \ref{EPS}, and show that in each region, one has a phase transition at the thick line in Figure 1 which is similar in nature to our transition between thermal AdS dominance and the black hole regime.

\begin{figure}\begin{center}
\label{fig:fundamental_domains}
\includegraphics[width=0.60\textwidth]{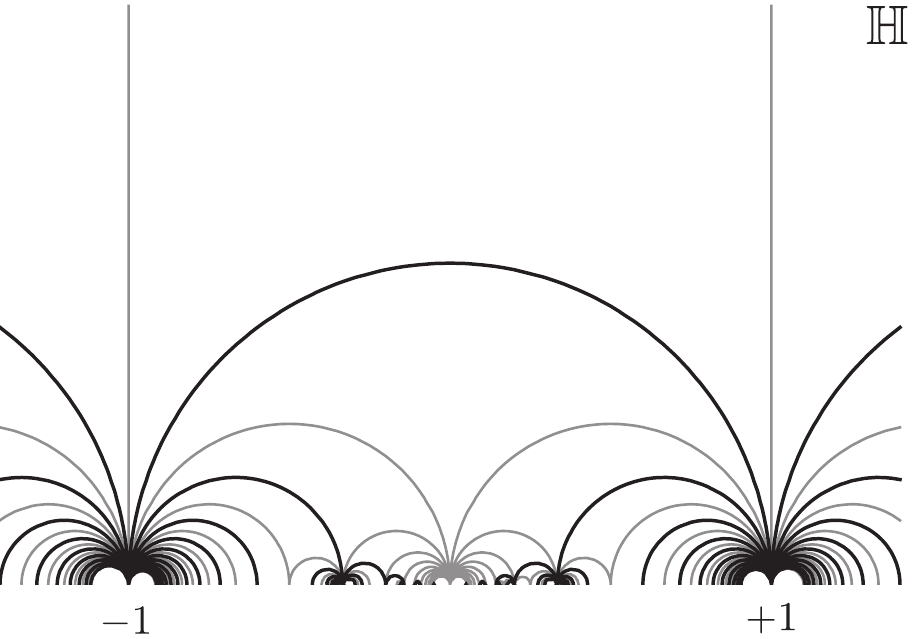}
\vspace{-10pt}
\caption{The tessellation by $\Gamma_\theta$ and its sub-tessellation by $\Gamma_\infty\backslash\Gamma_\theta$. The thick lines are where phase transitions in supergravity can occur. }
\end{center}
\end{figure}

\section{Examples}

In this section, we discuss how the elliptic genera of various simple CFTs -- $\sigma$-models with targets
${\rm Sym}^N(K3)$, product manifolds $(K3)^N$, or Calabi-Yau hypersurfaces up to relatively high dimension $d$ -- fare against the bound.  Somewhat unsurprisingly, the first class of theories passes the bound while the others fail dramatically, exhibiting far too rapid a growth in polar coefficients [\onlinecite{Keller:2011xi}].
We close with a discussion of ${\rm Sym}^{N}({\cal M})$, with ${\cal M}$ the Monster CFT of
Frenkel-Lepowsky-Meurman.  This example proves a useful foil in contrasting theories with
low energy supergravity vs low energy string duals.

\subsection{${\rm Sym}^N(K3)$}

The first example is one which we expect to satisfy the bound, and serves as a test of the bound.  A system
which historically played an important role in the development of the AdS/CFT correspondence was the D1-D5
system on $K3$ [\onlinecite{Strominger-Vafa}], and the duality between the $\sigma$-model with target space
$(K3)^N/S_N$ and supergravity in AdS$_3$ was one of the first examples of AdS$_3$/CFT$_2$ duality [\onlinecite{Maldacena}]. See also [\onlinecite{Dijkgraaf:1998gf}] for a more detailed analysis.

The elliptic genus of the symmetric product CFT was discussed extensively in [\onlinecite{DMVV}].  One can define a generating function for elliptic genera
\begin{equation}
\mc{Z}_X(\sigma,\t,z) = \sum_{N \geq 0} p^N Z_{R,R}(\text{Sym}^N(X);\tau,z)~,\quad p=e^{2\pi i \sigma},
\end{equation}
which is given by [\onlinecite{DMVV}] as
\be
\mc{Z}_X(\sigma,\t,z) = \prod_{n>0, n'\geq 0, l} \frac1{(1-p^nq^{n'}y^l)^{c_X(nn',l)}}.
\label{eq:dmvvgen}
\ee
The coefficients $c_X(n,l)$ are defined as the Fourier coefficients of the original CFT $X$,
\be
Z_{R,R}(X;\tau,z) = \sum_{n\geq0,l} c_X(n,l) q^ny^l.
\ee

If we are interested in calculating the $O(q^0)$ piece of the elliptic genus of ${\rm Sym}^N(X)$, we can set $n'=0$ in (\ref{eq:dmvvgen}), giving
\begin{equation}\label{eq:dmvvq0}
\lim_{\tau \to i\infty}\mc{Z}_X(\sigma,\t,z) = \prod_{n>0, l} \frac1{(1-p^n y^l)^{c_X(0,l)}} .
\end{equation}
When $X$ is the sigma model with Calabi-Yau target space (which we also call $X$), the above is, up to simple factors, the generating function for the $\chi_y$-genus of ${\rm Sym}^N(X)$.

The most polar term of ${\rm Sym}^N(X)$ is given by $y^{m N}$ where $m={\rm dim}_{\mathbb C}X/2$ is the index of the elliptic genus of $X$. This is the coefficient of $y^{m N} p^N$ in (\ref{eq:dmvvgen}), which only receives contributions from
\be
\frac1{(1-p y^{m})^{c_X(0,m)}}.
\label{eq:mostpolarpiece}
\ee
By calculating the coefficient of $p^N y^{N m}$ in (\ref{eq:mostpolarpiece}) we get
\be
c_{{\rm Sym}^N X}(0,Nm) =\binom{c_X(0,m) + N -1}{c_X(0,m)-1},
\ee
a polynomial of degree $c_X(0,m)-1$ in $N$ and therefore allowed by the bound (\ref{thebound}).

In order to find the subleading polar piece for ${\rm Sym}^N(X)$, we calculate the coefficient of the term $p^N y^{N m - 1}$ in (\ref{eq:dmvvgen}). This has contributions from
\be
\frac1{(1-p y^{m})^{c_X(0,m)}} \frac1{(1-p y^{m-1})^{c_X(0,m-1)}} \frac1{(1-p^2 y^{m})^{c_X(0,m)}}.
\label{eq:nextmostpolarpiece}
\ee

The $p^N y^{m N - 1}$ term generically comes from multiplying a $p^{N-1} y^{m (N - 1)}$ in the first term in (\ref{eq:nextmostpolarpiece}) with a $p y^{m-1}$ from the second term. For the special case of $m=1$, it can also come from multiplying a $p^{N-2} y^{m (N-2)}$ from the first term with a $p^2 y^{m}$ from the third term.

The coefficient of $p^{N-1} y^{m (N - 1)}$ in the first term is $\binom{c_X(0, m) + N - 2}{c_X(0,m)-1}$, and the coefficient of $p y^{m-1}$ in the second term is $c_X(0,m-1)$. The coefficient of $p^{N-2} y^{m (N-2)}$ in the first term is $\binom{c_X(0,m) + N - 3}{c_X(0,m)-1}$ and the coefficient of $p^2 y^{m}$ in the third term is $c_X(0,m)$. Thus the coefficient of the penultimate polar piece is given by
\be
c_{{\rm Sym}^N X}(0,Nm-1) =
\begin{cases}
\binom{c_X(0,m)+N-2}{c_X(0,m)-1} c_X(0,m-1),& \text{if } m>1\\
\binom{c_X(0,1)+N-2}{c_X(0,1)-1} c_X(0,0) + \binom{c_X(0,1)+N-3}{c_X(0,1)-1} c_X(0,1), & \text{if } m=1.
\end{cases}
\label{eq:mostgeneral}
\ee
Again, this exhibits polynomial growth in $N$ and is allowed by (\ref{thebound}). Any term a finite distance away from the most polar term (e.g. $y^{Nm-x} q^0$ for constant $x$) will grow as a polynomial in $N$ of degree $c_X(0,m)-1$.

For Calabi-Yau manifolds $X$ with $\chi_0 = 2$, we have $c_X(0, m)=2$ so the two most polar terms simplify to
\begin{align}
c_{{\rm Sym}^N X}(0,Nm) &= N +1 \nn \\
c_{{\rm Sym}^N X}(0,Nm-1) &=
\begin{cases}
N c_X(0,m-1), &\text{if } m>1\\
N c_X(0,0) + 2(N-1), &\text{if } m=1.
\end{cases}
\label{eq:lessgeneral}
\end{align}

For the special case of $X = K3$, we have $m=1$ and $c_X(0,0) = 20$, so the penultimate polar piece grows as $22N-2$.

We can do a similar calculation to find the coefficient in front of $y^{N-x}$ for $\text{Sym}^N(K3)$ with $x>1$. We find the asymptotic large $N$ value for the coefficient, presented in Table \ref{table:symn}.
In Figure 2, we plot the polar coefficients of ${\rm Sym}^{20}(K3)$ against the values allowed
by the bound. Although some very polar terms exceed $e^{2\pi(n - \frac{|\ell|}{2} + \frac{m}2)}$ in (\ref{thebound}), the deviation is of the order $O(\log{N})$ in the exponent, which is allowed in our analysis. For terms with polarity close to zero, the $O(\log{N})$ corrections are less important, and we see that the bound is subsaturated as expected.

\begin{table}[ht]
\caption{Coefficient of $y^{N-x}$ in $\text{Sym}^N(K3)$ elliptic genus at large $N$. We later plot these values in Figure 8.}
\centering
\begin{tabular}{c c c c}
$x$ & Coefficient\\
\hline
0 & $N+1$\\
1 & $22N-2$\\
2 & $277N-323$\\
3 & $2576N-5752$\\
4 & $19574N-64474$\\
5 & $128156N-557524$\\
6 & $746858N-4035502$\\
7 & $3959312N-25550800$\\
8 & $19391303N-145452673$\\
9 & $88757346N-758554926$\\
10 & $383059875N-3673549725$\\
11 & $1569800280N-16690133400$\\
12 & $6143337474N-71708443374$\\
13 & $23066290212N-293213888652$\\
14 & $83418524934N-1146991810674$\\
15 & $291538891984N-4310932524176$\\
16 & $987440609467N-15624074962373$\\
17 & $3249156243514N-54773846935526$\\
18 & $10408875430635N-186236541847125$\\
19 & $32525691116400N-615565850482800$\\
20 & $99302600734650N-1981904206578750$\\
\hline
\end{tabular}
\label{table:symn}
\end{table}

\begin{figure}
\begin{center}
\includegraphics[width=0.70\textwidth]{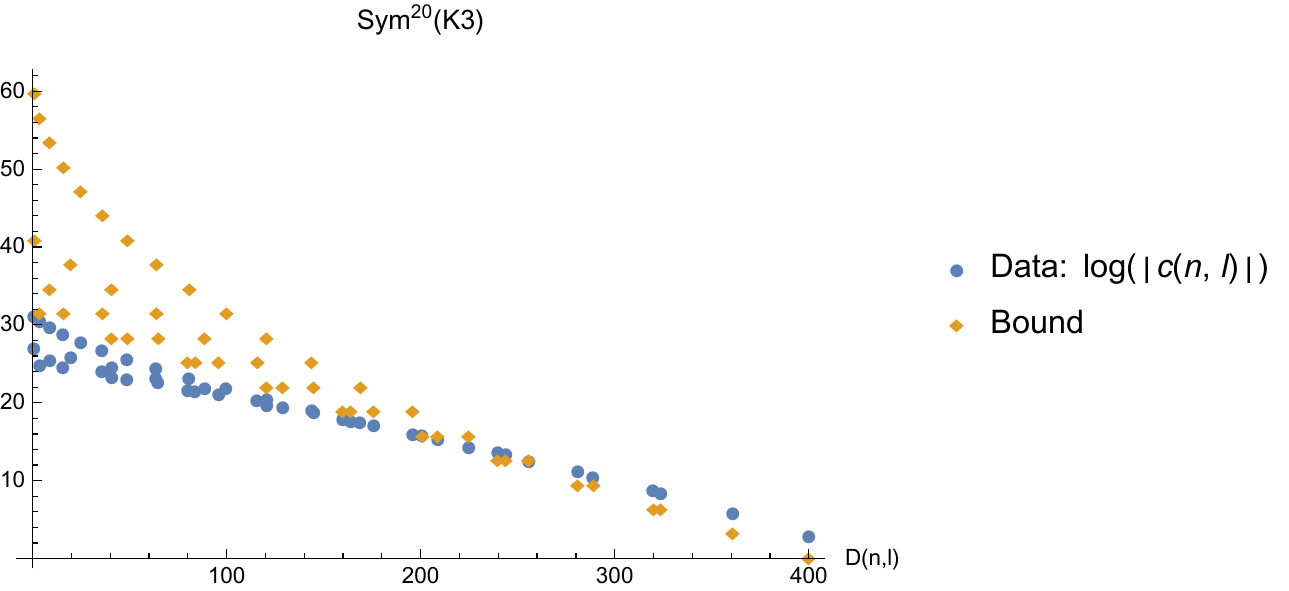}
\end{center}
\caption{
Here, we plot the polar coefficients of ${\rm Sym}^{20}(K3)$ versus polarity, and also the coefficients allowed by the bounds.  We see that at this value of $c$ (=120), the bounds are satisfied by the symmetric product conformal field theory, after allowing minor shifts due to the $O(\log{m})$ correction.}
\end{figure}

The fact that ${\rm Sym}^N (K3)$ satisfies our bounds is part of a more general story -- in fact all symmetric products will satisfy this bound, regardless of the ``seed" SCFT.   This follows from the general class of arguments presented in  [\onlinecite{Hartman}, \onlinecite{Keller:2011xi}].

\subsection{Products of $K3$ (or, $X^N$)}

The most obvious families of CFTs that ``should" fail any reasonable test for having a
(weakly coupled) gravity dual are given by tensor products of many small $c$ CFTs.  Here, as
a foil to $\text{Sym}^{N}(K3)$, we describe the results for the product $(K3)^N$.  Not surprisingly, it fails to
satisfy the bounds. We will use the fact that
\begin{align}
Z_{R,R}^{(X^{\otimes N})}(\t,z) = \big(Z_{R,R}^{X}(\t,z)\big)^N = \big(\sum_{n,\ell} c_X(n,\ell) q^n y^\ell \big)^N.
\end{align}

For concreteness, we look at the $\chi_y$ genus of $K3^N$.
Since
\begin{equation}
Z_{R,R}^{(K3)}(\t,z) = 2y^{-1} + 20 + 2y + O(q),
\label{eq:k3blahblahblah}
\end{equation}
the $q^0 y^N$ term in the elliptic genus of $K3^N$ is given by
\be
c_{K3^N}(0,N) = 2^N,
\ee
which violates the bound (\ref{thebound}) of only polynomial growth for the most polar term.

Actually we see that violations are prevalent. For instance, the $q^0 y^0$ term in the elliptic genus satisfies
\be
20^N < c_{K3^N}(0,0).
\ee
This is a clear underestimate of the actual coefficient, obtained by simply ignoring the $y^{-1}$ and $y$ terms in (\ref{eq:k3blahblahblah}).

However, even the underestimate violates the bound (\ref{thebound}) of
\be
c_{K3^N}(0,0) < e^{\frac{\pi}{2}N + O(\log N)} \sim 4.8^{N + O(\log{N})}.
\ee
To visualize the violation we plot the polar coefficients of $K3^{20}$ against the bound in Figure~3. Note that the violations are not of the order $O(\log N)$, and (\ref{thebound}) is clearly not satisfied.

\begin{figure}
\begin{center}
\includegraphics[width=0.70\textwidth]{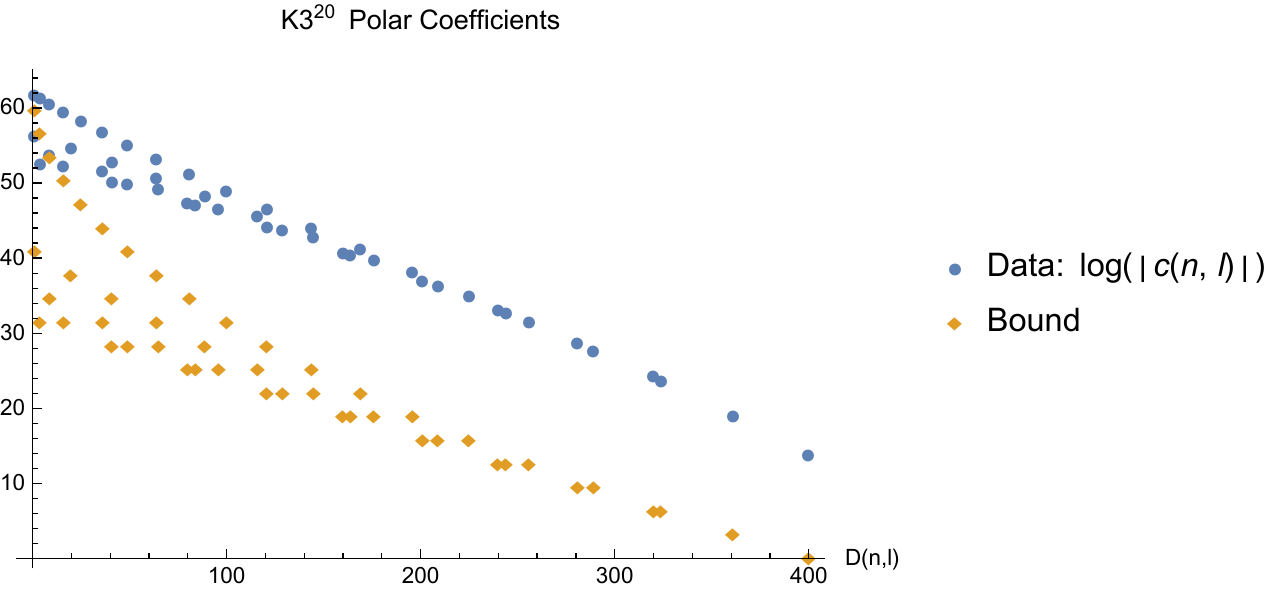}
\end{center}
\caption{
Here, we plot the polar coefficients of the product conformal field theory with target $K3^{20}$.}
\end{figure}

We conclude with a few remarks about examples similar to the above:

\begin{enumerate}

\item We cannot rule out \emph{all} product manifolds using this method. For instance, the elliptic genus of $T^4$ is zero, which means that products of $T^4$ will surely satisfy the bound, having a vanishing elliptic genus.
The vanishing is due to cancellations arising from the $U(1)^4$ translation symmetry acting on ${\rm Sym}^N(T^4)$.
One could instead work with ${\rm Sym}^N(T^4)/T^4$. In worldsheet terms, there are fermion zero modes due to
the extra translation symmetry which must be saturated by the insertion of a suitable number of fermion currents.
The relevant modification of the genus is worked out in  [\onlinecite{Gregtorus}].
It should be fairly straightforward to generalize our considerations to situations such as this where
extra insertions are required to define a proper index.

\item Another simple example that violates the bound is the iterated symmetric product
${\rm Sym}^{N_1} ({\rm Sym}^{N_2}(K3))$.  Taking, for simplicity, $N_1=N_2 = N$, so $m=N^2$,  the
coefficient of the most polar term is ${2N \choose N } \sim \frac{1}{\sqrt{\pi N}} 4^N = \frac{1}{\pi^{1/2} m^{1/4}} 4^{\sqrt{m}}$ for large $m$.
Indeed, the iterated symmetric product is an example of the more general class of
permutation orbifolds. It would be interesting to explore the relation of our
bound to the oligomorphic criterion of  [\onlinecite{Haehl:2014yla}, \onlinecite{Belin:2014fna}].

\end{enumerate}

\subsection{Calabi-Yau spaces of high dimension}

To provide a slightly more nontrivial test, we discuss the elliptic genera of Calabi-Yau sigma models with target spaces $X^{(d)}$
given by the hypersurfaces of degree $d+2$ in $\mathbb{CP}^{d+1}$, e.g.
\begin{equation}
\sum_{i=0}^{d+1} z_i^{d+2} ~=~0.
\end{equation}
We have chosen these as the simplest representatives among Calabi-Yau manifolds of dimension $d$;
as they are not expected to have any particularly special property uniformly with dimension, we suspect this
choice is more or less representative of the results we could obtain by surveying a richer class of
Calabi-Yau manifolds
at each $d$.  In any case we will settle with one Calabi-Yau per complex dimension. Since $m=d/2$, and we
have been assuming $m$ is integral, we restrict to even $d$.

The elliptic genus for these spaces is independent of moduli, and can be conveniently computed in the Landau-Ginzburg orbifold phase.  This yields the formula [\onlinecite{Yang}]
\begin{equation}
Z_{R,R}^{d}(\t,z) = \frac{1}{d+2}\sum_{k, \ell=0}^{d+1} y^{-\ell} \frac{\theta_1\(\t, -\frac{d+1}{d+2}z + \frac{\ell}{d+2} \t + \frac{k}{d+2}\)}{\theta_1\(\t, \frac{1}{d+2} z + \frac{\ell}{d+2} \t + \frac{k}{d+2}\)}
\end{equation}
Many further facts about elliptic genera of Calabi-Yau spaces can be found in [\onlinecite{Gritsenko}].

  First, we discuss the explicit data. To facilitate this we computed all polar coefficients numerically for $d=2,4,...,36$.  Then, we provide a simple analytical proof of bound violation valid
for all values of $d$ (just following from the behavior of the subleading polar term).

In Figures 4, 5, and 6 we plot the coefficients of the polar pieces against polarity for Calabi-Yau 10-, 20-, and 36-folds,
respectively.  In Figure 7, we plot the subleading polar coefficients of these Calabi-Yau spaces as a function of their dimension.
In all cases, we see that the bounds are badly violated.

\begin{figure}
\label{cy10}
\begin{center}
\includegraphics[width=0.70\textwidth]{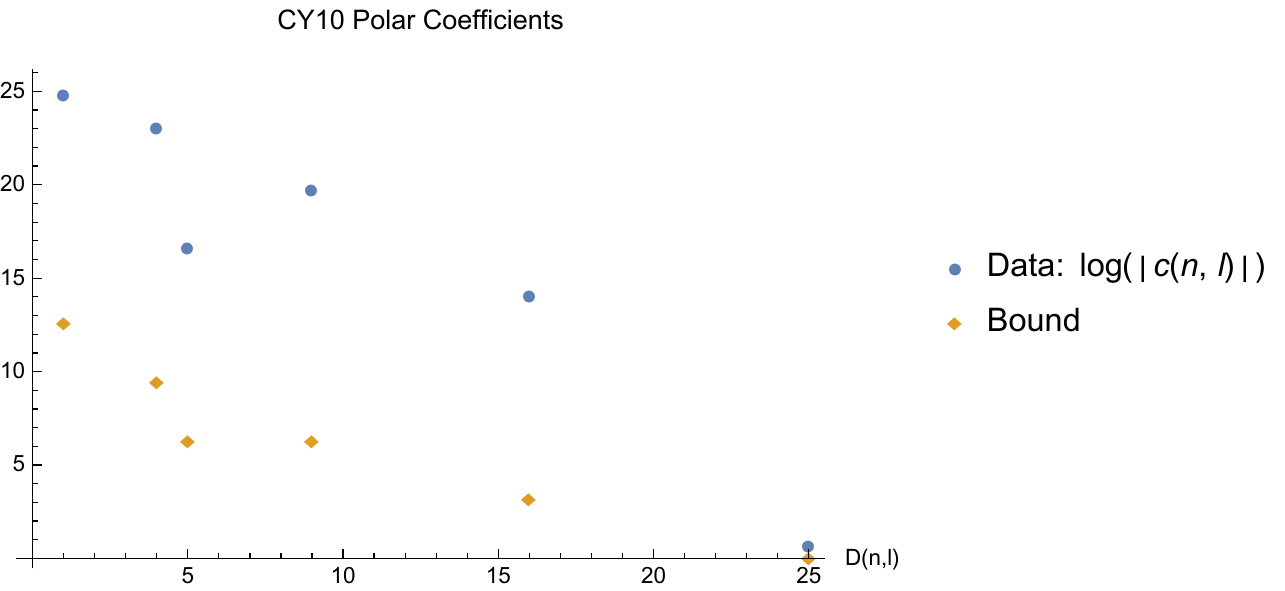}
\end{center}
\caption{
Here, we plot the polar coefficients of $Z_{RR}^{d=10}$.}

\label{fig:cy10}
\end{figure}

\begin{figure}
\label{cy20}
\begin{center}
\includegraphics[width=0.70\textwidth]{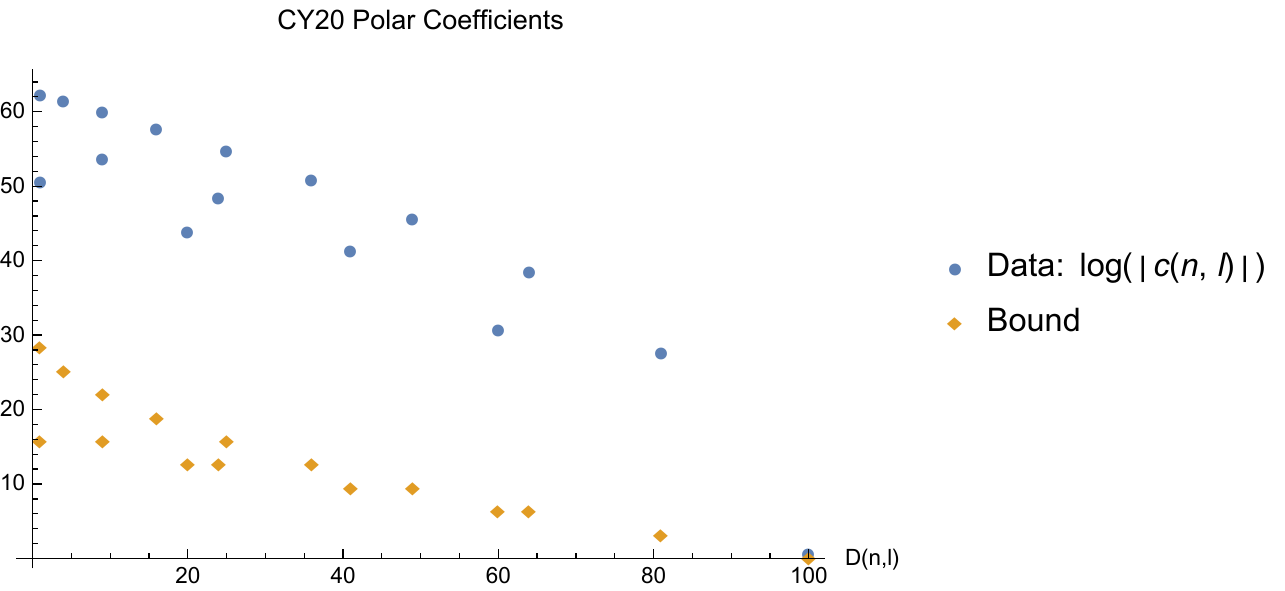}
\end{center}
\caption{
Here, we plot the polar coefficients of $Z_{RR}^{d=20}$.}

\label{fig:cy20}
\end{figure}

\begin{figure}
\label{cy36}
\begin{center}
\includegraphics[width=0.70\textwidth]{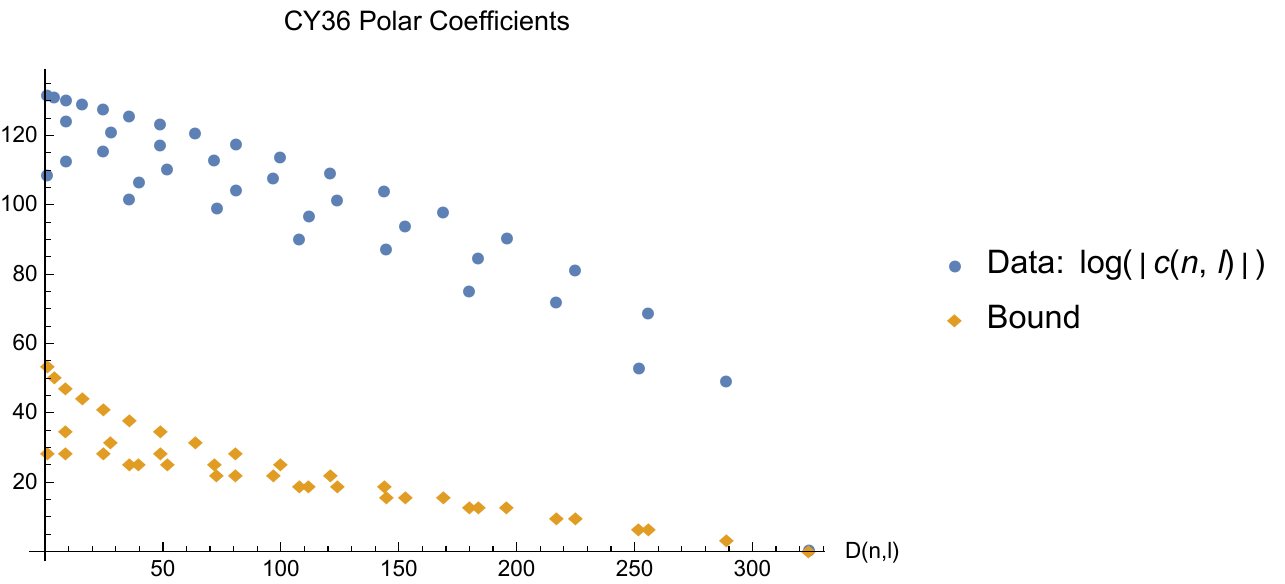}
\end{center}
\caption{
Here, we plot the polar coefficients of $Z_{RR}^{d=36}$.}

\label{fig:cy36}
\end{figure}

\begin{figure}
\label{cysub}
\begin{center}
\includegraphics[width=0.65\textwidth]{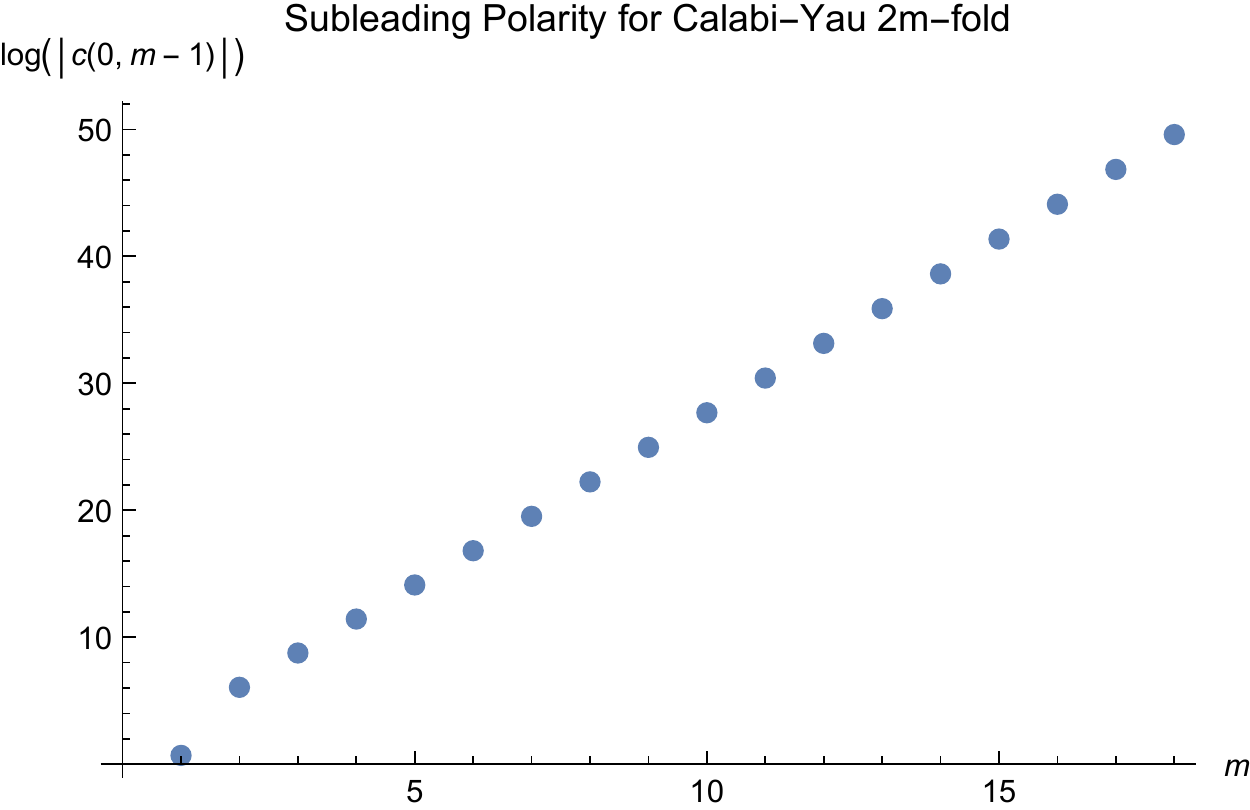}
\end{center}
\caption{
Here, we plot the subleading polar coefficients of the Calabi-Yau elliptic genera against the dimension.}
\label{fig:cysub}
\end{figure}

Numerics aside, it is easy to give a simple analytical argument proving that these Calabi-Yaus will violate the bound.
Consider the subleading $y^{m-1}$ polar piece of $Z_{RR}^{d=2m}$.

The coefficients $c_{X^{(d)}}(0,p)$ of the elliptic genera of Calabi-Yau spaces are determined simply by topological invariants:
\begin{equation}
\label{topstuff}
c_{X^{(d)}}(0,m-i) = \sum_k (-1)^{i+k} h^{k,i},
\end{equation}
so the coefficient in front of $y^{m-1}$ is
\be
-\chi_1 = \sum_p -(-1)^p h^{1,p}.
\ee
We know $h^{1,d-1}$ is given by the number of complex structure parameters of the hypersurface, or
\begin{align}
h^{1,d-1} &= \frac{(d+2)\times(d+3)\times\ldots\times(2d+3)}{1\times2\times\ldots\times(d+2)} -(d+2)^2 \nn \\
&={2d+3\choose d+2} -(d+2)^2.
\end{align}
By a standard application of the Lefschetz hyperplane theorem, the remaining $h^{1,p}$   vanish except for $h^{1,1} = 1$.
Thus we get (recall $d=2m$ is even)
\be
c_{X^{(d)}}(0,m-1) = {2d+3\choose d+2}-(d+2)^2+1.
\label{eq:sub}
\ee

And just as a check, for $d=36$, we numerically get
\begin{align}
c_{X^{(36)}}(0,17) &= 3446310324346630675857 \nn \\
&= {75\choose38} - 38^2 + 1
\end{align}
-- which matches the expectation on the nose.

Asymptotically, (\ref{eq:sub}) goes as:
\begin{align}
\log{c_{X^{(d)}}(0,m-1)} &\sim \log{(2d)!}-2\log{(d)!} \nn \\
&\sim 2d\log{(2d)} - 2d\log{(d)} \nn \\
&=2d\log{2}
\end{align}
so
\be
\label{ouch}
c_{X^{(d)}}(0,m-1)\sim2^{2d}=2^{4m}.
\ee

To satisfy the bound, we need $c_X^{(d)}(0,{m-1})$ to grow at most polynomially with $m$ when it in fact grows exponentially with $m$.

\subsection{Enter the Monster}

We now discuss a theory which passes our bounds but seemingly exhibits no supergravity regime --
instead exhibiting a Hagedorn degeneracy of states already at low energies.
We have benefited immensely in thinking about this theory from the unpublished work of Xi Yin.

A $c=24$ CFT with Monster symmetry was constructed many years ago by Frenkel, Lepowsky, and
Meurman [\onlinecite{FLM}].  Let us call the non-chiral CFT with Monster symmetry ${\cal M}$.  In this
section, we wish to consider the symmetric products ${\rm Sym}^N ({\cal M})$.  As ${\cal M}$ has
no moduli, there is a unique partition function canonically associated with this theory, and
we will consider the chiral partition function instead of the elliptic genus in this section.

This requires a word of explanation.  While the elliptic genera we've considered are related to non-chiral CFTs with conventional AdS gravity duals (in favorable cases), a chiral CFT can never have a conventional Einstein gravity dual.  However, as explained in [\onlinecite{Andyone}, \onlinecite{Andytwo}], there are candidates for chiral gravity duals to holomorphic CFTs. See also [\onlinecite{Skenderis:2009nt}] and references therein for a more detailed discussion on these  theories.
 In this sense, we can consider the partition functions which follow as (candidate) duals to (a suitably defined theory of) chiral gravity (coupled to suitable matter).

Using the formula for the second-quantized partition function [\onlinecite{DMVV}], along with
the famous denominator identity due to Borcherds [\onlinecite{Borcherds}]:
\be
\prod_{n>0,m\in \IZ} (1- p^n q^m)^{c(nm)} = p (J(\sigma) - J(\tau) )
\ee
where $p=e^{2\pi i \sigma}$ and $q = e^{2\pi i \tau}$ and $J(\tau) = q^{-1} + \sum_{n=1}^\infty c(n)q^n$,
 one can write the generating function:
\begin{equation}\label{eq:MonsterGenSum}
\sum_{N=0}^{\infty} e^{2\pi i N \sigma} Z({\rm Sym}^{N}({\cal M});\tau) ~=~{e^{-2\pi i \sigma} \over
{J(\sigma) - J(\tau)}}~.
\end{equation}
For large $\text{Im}(\tau)$ the infinite sum only converges for $\text{Im}(\sigma) > \text{Im}(\tau)$, while for small $\text{Im}(\tau)$ the
infinite sum only converges for $\text{Im}(\sigma+{1\over\tau}) > 1 $. Choosing large $\text{Im}(\tau)$ we can say that
\begin{equation}
Z({\rm Sym}^N({\cal M});\tau) ~=~\oint~d\sigma~{e^{-2\pi i (N+1) \sigma} \over {J(\sigma) - J(\tau)}}~.
\end{equation}
where the contour is a circle at constant $\text{Im}(\sigma)$ on the cylinder given by the quotient of the $\sigma$-plane by
$\sigma \sim \sigma +1$ and we must assume $\text{Im}(\sigma)>\text{Im}(\tau)$.
The contour integral can - at least naively -  be evaluated by deforming the contour to smaller values of $\text{Im}(\sigma)$
approaching $\text{Im}(\sigma)=0$. (We certainly cannot deform to large $\text{Im}(\sigma)$ because of the
exponential growth from the term  $e^{-2\pi i (N+1) \sigma} $.)  This deformation leads to residues from an
infinite set of simple poles at $\sigma = \tau $ together with $\sigma$ equal to
 all the modular images of $\tau$ within the strip $\vert \text{Re}(\tau)\vert \leq \frac{1}{2}$.  Using
\be
\frac{1}{2\pi i}\frac{\partial}{\partial \tau}J(\tau) = -\frac{E_4^2(\tau)E_6(\tau)}{\eta(\tau)^{24}},
\ee
this naive contour deformation yields:
\begin{equation}
Z({\rm Sym}^N({\cal M});\tau) ~=~{\bf P}_2(q^{-N-1}) {\eta(\tau)^{24} \over E_{4}(\tau)^2 E_6(\tau)}~.
\end{equation}
Here, ${\bf P}_2(q^{-N-1})$ is the weight 2 Poincar\'e series of $q^{-N-1}$.\footnote{This Poincar\'e series requires
regularization, indicating the above contour deformation argument is subtle.
A standard procedure for obtaining a well-defined Poincar\'e series
 is described in detail in many places. See, for examples,   Section 4
of [\onlinecite{ModernFaeryTail}] or Section 2 of [\onlinecite{ChengDuncan}].
As explained in those references, the modular anomaly of the series ${\bf P}_2(q^{-N-1}) $ is
expressed in terms of a period of a weight zero cusp form. Since no such nonzero cusp form exists we conclude that
${\bf P}_2(q^{-N-1})$ is in fact modular, as is required by
modularity of $Z({\rm Sym}^N({\cal M});\tau)$.}

Because
\begin{equation}
{\bf P}_2(q^{-N-1}) = q^{-N-1} + {\cal O}(1)~,
\end{equation}
all of the modes which provide the low-energy spectrum (i.e., the states which are not black holes)
are visible in the expansion of
\begin{equation}
F(\tau) = {\eta(\tau)^{24} \over E_4(\tau)^2 E_6(\tau)}~.
\end{equation}

It now follows from the fact that $c/24 = N$ and the structure of ${\bf P}_2$ that we
can find the modes at energies below the black hole bound just from expanding $F$.
Writing
\begin{equation}
F(\tau) = \sum_{n=1}^{\infty} a_n q^n~,
\end{equation}
$a_1$ is the ground-state contribution and the higher $a_k$ count the excited states
visible in the partition function (until one reaches the threshold to form black holes).

One can extract the $k^{\text{th}}$ coefficient via the contour integral
\begin{equation}
\label{explicit}
a_k ~=~{1\over 2\pi i} \oint ~d\tau~{1\over q^{k+1}} F(\tau)~.
\end{equation}
As $\eta(\tau)$ has no poles, $E_4$ has a simple zero at $\tau = e^{2\pi i \over 3}$ with no other zeroes, and $E_6$ has a simple zero
at $\tau = i$ with no other zeros, we can now evaluate (\ref{explicit}) explicitly.

The pole at $\tau=i$ provides the dominant behavior of the integral for $k \gg 1$.  One finds
\begin{equation}
a_k \sim e^{2\pi k} {\eta(i)^{24} \over E_4(i)^2 E_6^\prime(i)}~,
\end{equation}
and hence in the regime $1 \ll n \ll N = {c\over 24}$, the ${\rm Sym}^N({\cal M})$ theory has
a degeneracy of polar states governed by
\begin{equation}
\label{Hag}
a_n \sim e^{2\pi n}~.
\end{equation}

One can view this as satisfying an analog of the bound \eqref{thebound} for
chiral gravity. In harmony with this, the
singularity of \eqref{eq:MonsterGenSum} at $\sigma = \tau$ and at $\sigma = -1/\tau$ should come from the $N\to \infty$
limit of the partition functions, and this strongly suggests that the partition functions $Z({\rm Sym}^N(\CM);\tau)$
exhibit the expected Hawking--Page first order transition (as indeed follows from the general results
of [\onlinecite{Keller:2011xi}]), that is, the large $N$ asymptotics at fixed pure imaginary $\tau$ is given by:
\be
    Z({\rm Sym}^N({\cal M});\tau)
     \sim  \begin{cases} \kappa_1 N^{\kappa_2} q^{-N}(1+O(N^{-1}))   &    \text{Im}(\tau) \geq 1    \\
     \kappa_1 N^{\kappa_2} \tilde q^{-N}(1+O(N^{-1})) &    \text{Im}(\tau) \leq 1  \end{cases}
\ee
where $\tilde q := \exp(-2\pi i/\tau)$. Here $\kappa_1, \kappa_2$ are constants we have not attempted to determine.

The growth \eqref{Hag} exhibits a Hagedorn spectrum, hinting that if there is a holographically dual theory it
 must be a string theory with string scale comparable to the AdS radius.

\section{String versus Supergravity Duals}

We have just seen that some theories with a low-energy Hagedorn degeneracy
\begin{equation}
{\rm {\#} ~of~ states~ at~ energy~} n \sim e^{2\pi n},~~1 \ll n \ll {c\over 24}
\end{equation}
still satisfy our bounds. This might indicate that such theories are
 low-energy string theories -- there is no parametric separation
of scales evident between the emergence of a Hagedorn degeneracy and some other set of
low-energy modes with well-defined asymptotics (which could serve as a proxy for supergravity
KK modes)\footnote{Two subtleties could invalidate the considerations of this section.
In one direction, cancellations between terms in a partition function could lead
to subexponential growth of coefficients when in fact the entropy grows exponentially.
In the other direction, when considering the entropy at finite volume it can happen
that the entropy grows exponentially with energy, even though the theory is not a
string theory. For an example, see Section 7 of [\onlinecite{Galakhov:2013oja}].}.

This is to be contrasted with the growth of states exhibited by a supergravity theory in $d$
spatial dimensions, in the regime where the supergravity modes have wavelengths shorter than any
scale set by the curvature.  The gravity modes then behave, to
leading approximation, like a gas of free particles in $d$ dimensions.  The energy per unit volume
scales as
\begin{equation}
{\cal E} \sim T^{d+1}~,
\end{equation}
while the entropy per unit volume scales as
\begin{equation}
s \sim T^{d}~.
\end{equation}
Hence, in such a theory, one expects (simply from dimensional analysis) that
\begin{equation}
c_n \sim e^{{\rm const} \times n^\alpha},~~\alpha \equiv {d \over d+1}~
\end{equation}
in the regime dominated by supergravity modes.
For instance, in the canonical $AdS_5 \times S^5$ solution of IIB supergravity, there is a supergravity
regime with $E^{9 \over 10}$ growth of the entropy as a function of energy [\onlinecite{review}].

For $AdS_3 \times S^3 \times K3$ compactifications where the $K3$ is much
smaller than the $S^3$, one would expect a 6d supergravity regime to occur at low energies.
We now provide some simple analytical and numerical arguments demonstrating that the growth is indeed sub-Hagedorn.  Related discussions appear in [\onlinecite{Jan}, \onlinecite{Jantwo}].
The naive ``gas of particles" analogy discussed above, for polar terms, would suggest a growth of $e^{{\rm const} \times n^{5/6}}$.  One can get slower growth, however, due to cancellations in the supergravity modes which contribute to the elliptic genus.   We also note that at $g_{\rm string} \ll 1$, there would be a regime of energies in the full physical theory exhibiting a Hagedorn degeneracy of string states.  These do not, however, contribute in the elliptic genus.

First, we provide an analytical argument demonstrating that there is a range in which the polar terms of the elliptic genus of $\text{Sym}^N(K3)$ clearly has subexponential growth (though we do not quantify beyond this). Taking (\ref{eq:dmvvq0}) at $y=1$, we get that the sum of all $O(q^0)$ coefficients of the EG of $\text{Sym}^N(K3)$ is the $N^{\text{th}}$ coefficient of
\be
\frac{q}{\eta(\tau)^{24}}
\ee
which goes as
\be
e^{4\pi\sqrt{N}+ O(\log N)}.
\label{eq:twentyfourbosons}
\ee
Since all of the $O(q^0)$ pieces of the EG of $\text{Sym}^N(K3)$ are positive (which can be shown from (\ref{eq:dmvvq0}) for instance), each individual term must be smaller than (\ref{eq:twentyfourbosons}). If we label the $O(q^0)$ states by $n$ as above, we must have
\be
a_n < e^{4\pi \sqrt{N}}
\ee
Thus
\be
a_{N^{\alpha}} < e^{4\pi \sqrt{N}}
\ee
for $\alpha<1$ which correspond to states parametrically below the Planck mass in the NS sector as $N \rightarrow\infty$. Relabelling gives us
\be
a_n < e^{4\pi n^{\frac{1}{2\alpha}}}.
\label{eq:awesomegrowth}
\ee
We therefore find states parametrically lighter than the Planck mass with a subexponential growth of states. Note that there may be other states at the same energy level that we neglect due to only considering $O(q^0)$ terms in the elliptic genus. However, as we expect the entropy to be a function of polarity up to small corrections, taking terms with positive powers of $q$ into account would only multiply our expression in (\ref{eq:awesomegrowth}) by some polynomial factor without changing the leading order.

Because we expect the only relevant scales (other than supergravity KK scales) to be the string scale and Planck scale, and we do not get stringy growth in this regime, we expect subexponential growth throughout the polar terms.
We now provide further (weak) numerical evidence in favor of this hypothesis.
 We include  a plot of the normalized coefficients of $y^{N-x}$ for $x=1, \ldots 40$ in the large $N$ limit in Figure 8 (these numbers do not change past some $N$ since they only involve twisted sectors of permutations of some fixed length).

\begin{figure}
\begin{center}
\includegraphics[width=0.65\textwidth]{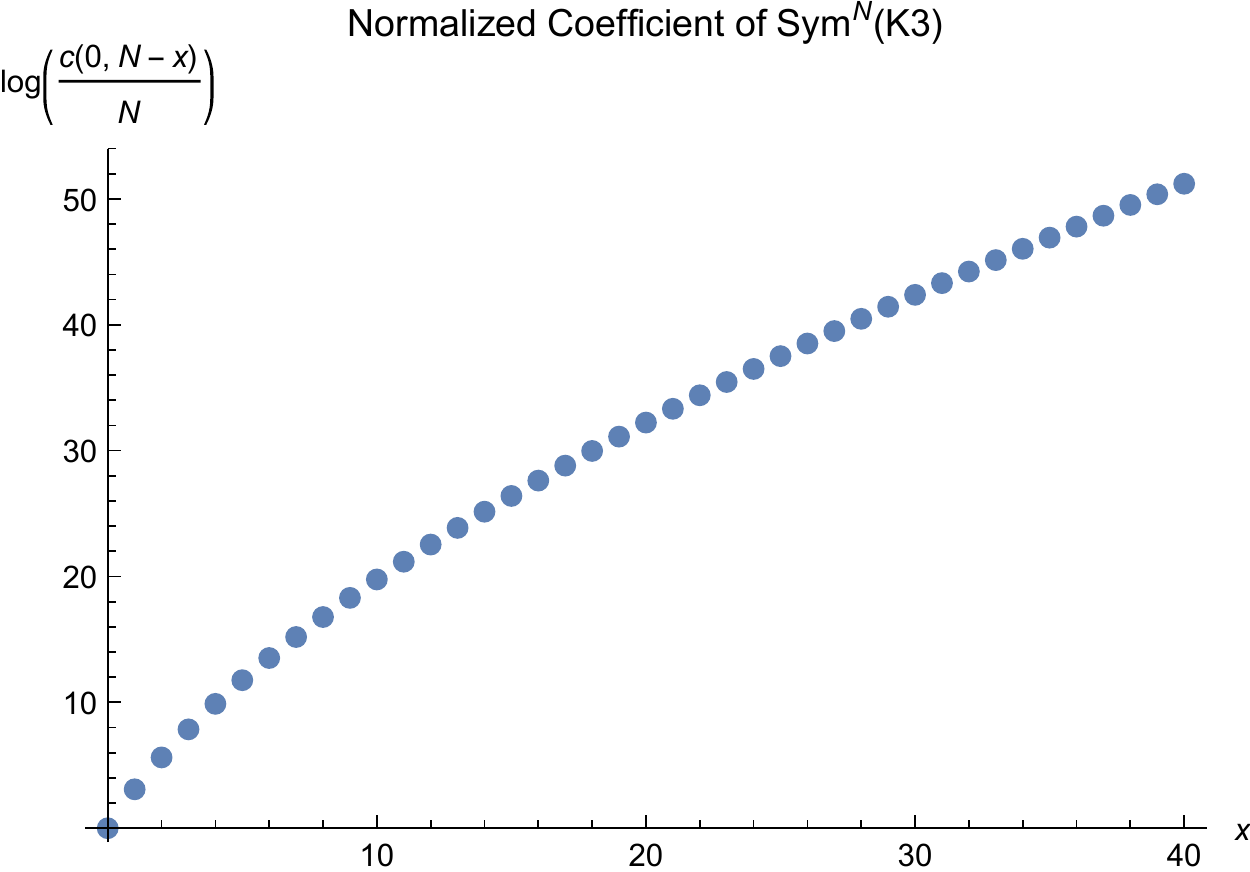}
\end{center}
\caption{
Here, we plot the normalized coefficients of $y^{N-x}$ terms in elliptic genus of $\text{Sym}^N(K3)$ for $x=1,\ldots 40$ in the large $N$ limit. Note the subexponential growth in the plot. Numerical values for the first twenty terms are given in Table I.}

\label{fig:symk3mostpolar}
\end{figure}

These examples suggest a criterion that distinguishes between theories with low-energy
Einstein gravity duals as opposed to low-energy string duals, with the usual qualifier that cancellation is possible in an index computation. Writing
\begin{equation}
c_n \sim e^{{\rm const} \times n^\alpha},~~1 \ll n \ll {c\over 24}~,
\end{equation}
theories with $\alpha < 1$ are likely to have a range of scales at low energy where supergravity
applies, while theories with $\alpha = 1$ are evidently string theories already at the scale set by the
curvature.  We note that similar issues have been discussed, in the context of the duality between
AdS$_4$ gravity and CFT$_3$, in the interesting paper [\onlinecite{Shiraz}].

\section{Estimating the volume of an interesting set of modular forms}
\label{sec:measure}

In this section we use (\ref{thebound}) to try and quantify a lower bound on the
``fraction of large $m$ superconformal field theories which may admit a gravity dual."
Our approach will be to ask:   ``How special is the class of weight zero, index $m$
Jacobi forms  corresponding to such superconformal theories?'' As we have seen,
 thermodynamic arguments constrain the growth of the polar coefficients
provided there is a physically reasonable gravitational dual, so the problem
reduces to quantifying ``what fraction'' of all possible polar coefficients
corresponds to the theories with gravitational duals.

Since the Jacobi form is completely determined by its
polar coefficients,     the map from CFTs to elliptic genera
can be viewed as a map from the space of $(0,2)$ field theories
to a subset  $\CE \subset \IZ^{j(m)}$. Now, there  is a natural metric on the moduli spaces
of conformal field theories, namely, the Zamolodchikov metric [\onlinecite{Zam}].
The moduli
space of such theories, with a fixed central charge $c$,
 is a union of connected components $\amalg_{\alpha} \CM^{(c)}_{\alpha}$.
It was suggested some time ago that, at least for the space of
$(2,2)$ superconformal theories,  the total Zamolodchikov volume of
$V^{(c)}:= \sum_{\alpha} \vol(\CM^{(c)}_{\alpha})$ should be finite.
This was based on physical arguments
[\onlinecite{Horne:1994mi}, \onlinecite{Douglas:2005hq}].
For the case of components arising from Calabi-Yau manifolds it has
been shown that indeed $\vol(\CM^{(c)}_{\alpha})$ is finite.
(See [\onlinecite{Douglas:2006zj}] and references therein for the
mathematical work on this subject.)  The finiteness of $V^{(c)}$ would allow us to
define a measure on the space of $(2,2)$ theories of a fixed central charge
and thereby to quantify statements of ``how often'' a property is exhibited
in a natural way. We will assume that $V^{(c)}$ is in fact finite\footnote{D. Friedan has proposed a mechanism by which such a probability
distribution might in fact be dynamically generated from more fundamental
principles [\onlinecite{Friedan:2002aa}, \onlinecite{Friedan:2002fe}].}.

Using the push-forward measure under the map to the polar coefficients of
elliptic genera we obtain a natural measure on the space $\CE$ of polar coefficients.
Unfortunately, our present state of knowledge of conformal field theory is
too primitive to evaluate this measure in great detail, but to illustrate the idea, and some
of the issues which will arise,  we will sketch two toy computations.

\def\fe{\mathfrak{e}}
For our first toy computation we consider the pushforward to a measure on $\IZ_+$ for the
absolute value of the extreme polar
coefficient of the elliptic genus. We denote this by
\be
\fe(\CC) = \vert c(0,m;\CC)\vert
\ee
for a $(2,2)$ CFT with $c=6m$.

Now $\fe$ is multiplicative on CFTs,
\be
\fe(\CC_1\times \CC_2) = \fe(\CC_1)\fe(\CC_2).
\ee
We would also like to say the same for the volumes:
\be\label{eq:Volume-Mult}
\vol(\CC_1\times \CC_2) {\buildrel ? \over  =} \vol(\CC_1)\vol(\CC_2)
\ee
but this is in general not the case. A simple counterexample is provided by
conformal field theories with toroidal target spaces. Nevertheless, for
ensembles such as theories based on \emph{generic} Calabi-Yau manifolds
the volume is multiplicative, because the relevant Hodge numbers are additive.
We will refer to an ensemble of CFT's for which \eqref{eq:Volume-Mult} holds
as a \emph{multiplicative ensemble} and here we restrict attention to such
ensembles. Extending our discussion beyond multiplicative ensembles is an
interesting, but potentially difficult, problem.

Given a multiplicative ensemble, let us say an $N=(2,2)$  CFT  $\CC$ is  \emph{prime} if it is not the product
of two such theories $\CC_1$ and $\CC_2$ each with positive central charge. Let $\CC(m,\alpha)$ denote the
distinct prime CFT's of central charge $c=6m$, with $\alpha = 1, \dots, f_m$. We expect $f_m$ to be finite,
but this is not necessary for our construction, so long as the relevant products below converge.
Denote the absolute value of the extreme polar coefficient, and the Zamolodchikov volume of $\CC(m,\alpha)$ by
$\fe(m,\alpha),v(m,\alpha)$, respectively.
Then the Zamolodchikov volume  ${\rm vol}(M)$  of theories of central charge $c=6M$ is determined from:
\begin{equation}
\prod_{m=1}^{\infty}\prod_{\alpha=1}^{f_m}  \frac{1}{1- v(m,\alpha) q^m} = 1 + \sum_{M=1}^\infty {\rm vol}(M) q^M.
\end{equation}
Similarly, we can write a generating function for the volume of the theories with
a fixed extreme polar coefficient. We assume that $\fe(m,\alpha)\not=0$ in our 
ensemble (thus excluding, for example, Calabi-Yau models with odd complex dimension) 
and form the generating function:
\begin{equation}
\prod_{m=1}^{\infty}\prod_{\alpha=1}^{f_m}  \frac{1}{1- v(m,\alpha) \fe(m,\alpha)^{-s} q^m }
=  1 + \sum_{M=1}^\infty \xi(s;M) q^M
\end{equation}
Then
\be\label{eq:XiFun}
\xi(s;M)= \sum_{\fe=1}^\infty \frac{\vol(\fe;M)}{\fe^s }
\ee
and the measure for the extreme polar coefficient is
\be\label{eq:RelMeas}
\mu(\fe;M):= \frac{\vol(\fe;M)}{\vol(M)}.
\ee

In order to make this slightly more concrete let us restrict even further
to the   the ensemble of $(4,4)$ theories generated by taking
products of the symmetric products of $K3$ sigma models, such as
\be\label{eq:ProdTheory}
\left( {\rm Sym}^1(K3) \right)^{n_1} \times \left( {\rm Sym}^2(K3) \right)^{n_2}\times
\cdots \left( {\rm Sym}^\ell(K3) \right)^{n_\ell}.
\ee
We will call this the \emph{K3-ensemble} and it is a multiplicative ensemble of CFT's.
In this ensemble the prime CFTs are simply the symmetric products ${\rm Sym}^n(K3)$.
For ${\rm Sym}^1(K3)$ the moduli space $\CM_1$ is the famous double quotient
\be\label{eq:MdSpc-1}
\CM_1 = O(\Gamma) \backslash O(4,20;\IR)/O(4) \times O(20)
\ee
with $\Gamma \cong {\rm II}^4 \oplus E_8 \oplus E_8$, while for $N>1$ the moduli space is
[\onlinecite{Dijkgraaf:1998gf},\onlinecite{Giveon:1998ns},\onlinecite{Seiberg:1999xz}]
\be\label{eq:MdSpc-2}
\CM_N = O(\Gamma') \backslash O(4,21;\IR)/O(4) \times O(21)
\ee
with $\Gamma'$ a lattice of signature $4,21$ determined in [\onlinecite{Seiberg:1999xz}].
The four ``extra moduli'' in \eqref{eq:MdSpc-2} compared to \eqref{eq:MdSpc-1} are due to
the blowup multiplet at the locus of $A_1$ singularities in ${\rm Sym}^N(K3)$ where
two points meet. All higher twist fields are irrelevant. The Zamolodchikov volume
$v_N$ of these moduli spaces is the same as the volume in the Haar measure. The Haar measure is
determined up to a single scale factor, and the relevant normalization can be
 fixed by a computation in conformal field
theory. Although the $\CM_N$ for $N>1$ are all diffeomorphic they are not isometric.
When comparing volumes for different values of $N$ with $N>1$ we must be careful
about the relative normalizations of the Zamolodchikov metric, and this can be determined
by the following argument:   If an exactly
marginal operator $\CO$ in the K3 sigma model perturbs a modulus $\mu \to \mu + \delta \mu$ then
the exactly marginal operator $\CO = \CO^{(1)}+\cdots + \CO^{(N)}$ in the ${\rm Sym}^N(K3)$ theory
perturbs the modulus $\mu$ by the same amount. (Correctly normalizing
the operator $\CO$ can be a confusing point. To see that our choice is the correct one,
note that it is similar to the fact that the energy-momentum tensor of the symmetric
product theory is given by $T = T^{(1)}+\cdots + T^{(N)}$.)  Therefore the Zamolodchikov metric on $\CM_N$
is $N/2$ times larger than that on $\CM_2$. Therefore we can say that there are positive constants $v,w$ such that
\be\label{eq:Vols}
v_N = \begin{cases} v & N=1 \\
N^{d} w & N > 1 \\
\end{cases}
\ee
where $d = 42  = \half \dim(O(4,21)/O(4)\times O(21))$.

The Zamolodchikov volume ${\rm vol}(M)$ of the   ensemble of models \eqref{eq:ProdTheory} is simply given by
\begin{equation}
\prod_{n=1}^{\infty} \frac{1}{1- v_n q^n} = 1 + \sum_{M=1}^\infty {\rm vol}(M) q^M
\end{equation}
%
Now, to get the measure for a fixed extreme polar term we noted above
that
\be
\fe(  {\rm Sym}^n(K3)) = n+1,
\ee
so the extreme polar
term of the elliptic genus of \eqref{eq:ProdTheory} is just the product:
\be\label{eq:Factor}
2^{n_1} 3^{n_2} \cdots (\ell+1)^{n_{\ell}}.
\ee
Therefore, our general formula specializes to
\be
\prod_{m=1}^{\infty} \frac{1}{1-v_{m} (m+1)^{-s}q^m} = 1 + \sum_{M=1}^\infty \xi(s;M) q^M
\ee
where $\xi(s;M)$ defines the conditional volume as in \eqref{eq:XiFun} and the measure
for the extreme polar term is given by \eqref{eq:RelMeas}, above.

%

Determining the numerical values of the constants $v,w$ used above is a
very interesting problem in number theory. This will be discussed in a separate paper,
along with some applications of the function $\xi(s;M)$ to the central issue of this
paper.\footnote{For further details, see \url{https://www.perimeterinstitute.ca/video-library/collection/mock-modularity-moonshine-and-string-theory}.}
It would also be very interesting to extend the above discussion to the ensemble of
all $(4,4)$ theories, but this looks quite challenging.
We would need to include  products with
${\rm Sym}^N(T^4)/T^4$. Moreover, we have omitted products with other $(4,4)$ models
constructable from permutation orbifolds, or from other compact hyperkahler
manifolds arising from moduli spaces of hyperholomorphic bundles on $K3$ and $T^4$.
And we have omitted the unknown unknowns since we do not know that every $(4,4)$
model can be realized geometrically.   Nevertheless, we expect some of the
basic features of the above discussion to survive better knowledge of the moduli space.


The above discussion is our first toy computation.
Given our poor knowledge of the moduli space of conformal field
theories we will resort to a second toy computation. We hope it proves instructive.
We enumerate the polar coefficients $c(a)$ by decreasing discriminant $D(a) = \ell(a)^2 - 4mn(a)$,
$a=0,\dots, j(m)-1$ where $j(m)= \dim \tilde J_{0,m}$. Thus, $D(0)=m^2$.
The idea of the second toy computation is to find a natural probability measure on
the vector space of   polar coefficients $(c(0), \dots, c(N))$. Of course,
a vector space has infinite measure in its Euclidean norm so we map these
coefficients to an  affine
coordinate patch of $\IR \IP^{N}$, with $N= j(m)$. That is, we consider the
points $[1:c(0):\cdots : c(N)]$ in $\IR \IP^{N}$. We then consider
the Fubini-Study measure on this patch.  Whether this measure
bears any relation to the \emph{a priori} Zamolodchikov measure (in the large $N$
limit) remains to be seen. (Since we do not like the answer, we suspect the answer is
that it does not.)

The volume element for the unit radius $\IR \IP^N$ in affine coordinates $[1:\xi^1:\cdots : \xi^N]$ is:
\be
d\vol =   \frac{ d \xi^1 \wedge \cdots \wedge d \xi^N }{(1+\sum_a   (\xi^a)^2 )^{(N+1)/2}}.
\ee
Now we consider the subspace of the affine coordinate patch with
\be
\vert c(a) \vert \leq R(a).
\ee
$R(a)$ is a bound which is supposed to come from physics. One reasonable
guess is
\be
R(a) = e^{2\pi (n(a) - \frac{|\ell(a)|}{2} + \frac{m}2)}.
\ee
Note that this is imposing (\ref{thebound}) without allowing an $O(\log m)$ correction.
Concretely, we are interested in the fraction
\be
f_N=\frac{\Gamma(\frac{N+1}2)}{\pi^{\frac{N+1}2}}\(\prod_{i=1}^N \int_{-e^{2\pi (n_i - \frac{|\ell_i|}{2}+\frac{m}2)}}^{e^{2\pi (n_i - \frac{|\ell_i|}{2}+\frac{m}2)}} d\xi_i\) \frac{1}{(1+\sum_i(\xi^i)^2)^{\frac{N+1}2}}
\ee
in the limit $N \rightarrow \infty$.

In Appendix \ref{estimathon}, we show that in the limit of large $N$,
\be
0.9699 < f_N < 0.9725.
\ee
We actually view this as a good indication that the Fubini-Study measure is \emph{not} a
good surrogate for the Zamolodchikov measure.
On general grounds, one actually expects theories
with weakly coupled gravity duals (even characterizing some small region of their moduli space) to be
rare creatures.

In general CFTs, the number of excited states at large energies $n$
grows like $e^{2\pi \sqrt{{c\over 6} n}}$ by the Cardy formula.  Hence a measure which was based on
``expecting" there to be a small number of states in that regime would clearly be incorrect.  While
one cannot use Cardy's result in the energy range characterizing polar coefficients, it seems suspicious
that our measure ``expects" the fewer polar coefficients -- related to states with high energy, though below the black hole bound -- to be close to 0.  In fact, one might expect that in a random SCFT,
the polar coefficients typically grow fairly rapidly with decreasing polarity.  In such a case, it would
be more difficult for them to lie within the polydisc specified by our bounds.  Finding a modified
volume estimate (or attaching a plausible physical meaning to our present estimate) will have to remain
a problem for the future.

\acknowledgments{
We thank J. de Boer, A. Castro, E. Dyer, A.L. Fitzpatrick, D. Friedan, S. Harrison, K. Jensen, C. Keller, S. Minwalla, D. Tong, S.P. Trivedi, E. Verlinde, R. Volpato, and X. Yin for interesting discussions.  We thank D. Ramirez for help with sophisticated mathematical diagrams and C. Keller for useful comments on the draft.
This work was initiated at the Aspen Center for Physics and we thank the ACP for its hospitality and excellent working conditions.
The ACP is supported by NSF Grant No. PHY-1066293.
N.B. acknowledges the support of the Stanford Graduate Fellowship.  N.M.P. acknowledges support from an NSF Graduate Research Fellowship.  S.K. is grateful to  the 2014 Indian Strings Conference, IISER-Pune,
the Tata Institute for Fundamental Research, and the Kavli Institute for Theoretical Physics at UC Santa Barbara for hospitality while thinking about the physics of this note.
His work was supported in part by National Science Foundation grant PHY-0756174 and DOE Office of Basic Energy Sciences contract DE-AC02-76SF00515.
M.C. is grateful to Cambridge, Case Western Reserve, and Stanford Universities and Max Planck Institut f\"ur Mathematik for hospitality. The work of   G.M. is supported by the DOE under grant
DOE-SC0010008 to Rutgers,   and NSF Focused Research Group
award DMS-1160461.   }

\appendix

\section{Extended phase diagram}
\label{EPS}

Here, we derive in detail the extended phase diagram depicted in Figure 1. The logic of the argument can be summarized as follows. The expression of the elliptic genus as a regularized Poincar\'e sum involves a sum over all co-prime pairs of integers $(c,d)$. For each such pair arising from the invariant group $\Gamma_\theta$ of $Z_{NS,+}$, we will find that there is an $(n,\ell)$ labelling a polar term in the elliptic genus which can serve as the analogue of our ground-state in the ground-state dominance condition. As a consequence, in each region in the tessellated upper half-plane there is a single pair $(c,d)$ labelling the saddle point which dominates the gravitational path integral (CFT elliptic genus). Each phase transition across the bold lines in Figure 1 is then a modular copy of the one we studied in this paper.

The elliptic genus can be written in terms of its polar part as  [\onlinecite{ModernFaeryTail}]
\begin{align}\notag
Z_{R,R}(\t,z) &= \frac{1}{2} \sum_{r \in \ZZ/2m\ZZ} C_r(0)\,\theta_{m,r}(\t,z) + \frac{1}{2}  \sum_{\substack{\ell\in \ZZ,n \geq0 \\   D(n,\ell)>0 }} \lim_{K\to\infty}\sum_{(\Gamma_\infty\backslash \Gamma)_K}\\\label{EG_polar}
&C_\ell(D(n,\ell))\exp\left(2\pi i\big(n{a\t+b\over{c\t+d}}+\ell {z\over{c\t+d}} -m{cz^2\over{c\t+d}}\big)\right)\,R\big({2\pi i D(n,\ell)\over 4m\,c(c\t+d)}\big)
\end{align}
where the limit coset is given by
\be
\lim_{K\to\infty}\sum_{(\Gamma_\infty\backslash \Gamma)_K} :=  \lim_{K\to\infty} \sum_{\substack{0<c< K\\ -K^2 < d <K^2 \\ (c,d)=1 }}
\ee
and $R$ is the regularization factor
\be
R(x) = \frac{2}{\sqrt{\pi}}\int_0^x e^{-z}z^{1/2} dz = {\rm erf}(\sqrt{x}) - 2 \sqrt{x\over \pi} e^{-x}
\ee
where ${\rm erf}(z)={2\over\sqrt{\pi}}\int_0^z e^{-t^2}dt$ denotes the error function.

As discussed in  [\onlinecite{FaeryTail}, \S 6], using the classic identities
\begin{align}\notag
{a\t+b\over c\t+d} &= {a\over c}- {1 \over{c(c\t+d)}}\\\notag
{1\over2}{\t+1\over c\t+d}&={1\over2c}-{d\over 2} {1 \over{c(c\t+d)}}+{1\over2}{1\over c\t+d}\\\notag
{c(\t/2+1/2)^2\over c\t+d} &={\t\over4}+{{2c-d}\over4c}+ {1\over 4}{{c^2 -2cd +d^2} \over{c(c\t+d)}}
\end{align}
and ${\rm Im}(-\frac{1}{c(c\t+d)})={ {\rm Im}(\t)  \over  |c\t+d|^2} =  {\rm Im}({a\t+b\over c\t+d}) $, we see that

\begin{align}
Z_{NS,+}(\tau) &= (-1)^m q^{m\over 4} \frac{1}{2} \sum_{r \in \ZZ/2m\ZZ} C_r(0)\,\theta_{m,r}\(\t,\frac{\t+1}{2}\) \nn\\ &+
 \frac{1}{2}  \sum_{\substack{\ell\in \ZZ,n \geq0 \\   D(n,\ell)>0 }} \lim_{K\to\infty}\sum_{(\Gamma_\infty\backslash \Gamma)_K} X(n,\ell;c,d)
 R\({2\pi i D(n,\ell)\over 4mc(c\t+d)}\)
\end{align}
with
\be\label{exponential}
\Big\lvert X(n,\ell;c,d) \Big\lvert  = |C_\ell(D(n,\ell))| \, \exp\left(-2\pi \,{\rm Im}\big({a\t+b\over c\t+d}\big) \,\big(m{(d-c)^2\over4} + n + \ell {d-c\over 2}\big)\right).
\ee

We would like to know which term in the elliptic genus, i.e. which pair $(n,\ell)$, contributes the most to the sum in (\ref{EG_polar}) with a given pair $(c,d)$.
First, focusing on the exponential factor in (\ref{exponential}), using that ${\rm Im}\big({a\t+b\over c\t+d}\big)>0$ and $0<D(n,\ell)\leq m^2$ we conclude that
the maximum of $\exp\left(-2\pi \,{\rm Im}\big({a\t+b\over c\t+d}\big) \,\big(m{(d-c)^2\over4} + n + \ell {(d-c)\over 2}\big)\right)$ occurs at $(n,\ell) =(n_{c,d},\ell_{c,d})$,
\[
(n_{c,d},\ell_{c,d}):= (\,\tfrac{m}{4} ((d-c)^2-1), -m(d-c))
 \] when $d-c$ is odd.
Ignoring the other factors for the moment, we expect that $\Big\lvert X(n,\ell;c,d) \Big\lvert$ has its maximum
\be\label{assume_thatitworks}
\Big\lvert X\big(n_{c,d},\ell_{c,d};c,d\big) \Big\lvert = C_{-m}(m^2) \,\exp\left( 2\pi \frac{m}{4}{\rm Im}\big({a\t+b\over c\t+d}\big) \right)
\ee
 when $(n,\ell) = (n_{c,d},\ell_{c,d})$. In the above we have used the fact that $c(n_{c,d},\ell_{c,d}) =c(0,m) = C_{-m}(m^2)$ is equal to the number of NS ground states (see (\ref{jac_coeff})).

The situation is different for the pair of co-primes integers $(c,d)$ with even $d-c$.
Using the more refined condition for the discriminants of the polar terms
$$0<D(n,\ell)\leq r^2 ~~ {\rm where} -m < r \leq m ,~ \ell=r\mod{2m}$$  that holds for all weak Jacobi forms as a straightforward consequence of (\ref{eqn:forms:jac:thetaxpn}), we see that the maximum of the exponential term
in (\ref{exponential}) is of order $1$ which is achieved whenever $\ell = -(d-c)m + r$, $n=m(d-c)^2 -\frac{(d-c)r}2$ for any $ -m < r \leq m$.
In other words, the contribution of the the part of the sum given by a pair $(c,d)$ with $c-d\equiv0\pmod{2}$ in (\ref{EG_polar}) is exponentially suppressed.

As a result, assuming that the exponential factor in (\ref{assume_thatitworks}) is the dominating factor and ignoring for the moment the regularization factor, one concludes
  that in each region in the upper-half plane given by the tessellation by $\Gamma_\infty\backslash \Gamma_\theta$ there is a unique pair $(c,d)$ that dominates and this corresponds to the infinitely many phases of 3d quantum gravity.
  To see this, notice that
 \[
{\rm Im}\big({a\t+b\over c\t+d}\big) ={ {\rm Im}(\t)\over|c\t+d|^2} \leq  {\rm Im}(\tau)\quad \forall \begin{pmatrix}a&b\\c&d\end{pmatrix}\in \Gamma_\theta
 \]
 whenever $\tau \in \Gamma_\infty {\cal F}$ is in the (interior of the) fundamental domain
 $${\cal F} = \{\tau \in \mathbb{H} \lvert |\tau| > 1 , ~-1<{\rm Re}(\tau)<1 \}
 $$ of $\Gamma_\theta$ or any of its images under the translation $\tau \to \tau +2n$, $n\in \ZZ$. See Figure 1.

Next we would like to discuss the conditions under which that the term with $(n,\ell)=(n_{c,d},\ell_{c,d})$ indeed dominates the sum over all polar terms for a given pair $(c,d)$. First we show that the effect of the regularization factor can be ignored at the large central charge limit where $D(n_{c,d},\ell_{c,d})/4m=m/4 \gg 1$.
To see this, note that $R(x) \to 0$ as $x\to 0$ and
\[
R(x)-1 = O(\sqrt{x}\,e^{-x})
\]
as $x\to \infty$, and
\[
{\rm Re}\big( \frac{2\pi i D(n,\ell)}{4mc(c\t+d)}\big) = \frac{2\pi D(n,\ell)}{4m}\,{\rm Im}\big({a\t+b\over c\t+d}\big).
\]
Second, for there to be no term over dominating the term coming from  $(n,\ell)=(n_{c,d},\ell_{c,d})$ in the sum in the region where ${{a\t+b\over c\t+d}\in\Gamma_\infty {\cal F}}$ as predicted by analysing the exponential factor alone as in (\ref{exponential}), focussing on the line ${\rm Re}({a\t+b\over c\t+d})=0$ we see that the coefficients of the polar terms have to satisfy
\be\label{sl2z_bound}
\log| c(n_{c,d},\ell_{c,d})|  \leq \left( 2 \pi  (\frac{m((d-c)^2+1)}{4} + n + \frac{\ell (d-c)}2) \right) + O(\log m)
\ee
for all co-prime pairs $(c,d)$ with $d-c$ odd.
It is not hard to show that the seemingly stronger condition (\ref{sl2z_bound}) is in fact implied by our bound (\ref{deltaonebound}) when taking the spectral flow symmetry into account. Recalling that $c(n_{c,d},\ell_{c,d})=C_{-m}(m^2)$ and
$$c(n,\ell) = c(n(k),\ell(k))~,~ n(k)=n+k^2m + k\ell, ~\ell(k)= \ell+2k m$$
 for all $k\in \ZZ$, we can write (\ref{sl2z_bound}) as
\[
\log | {c(n(k),\ell(k))}  |\leq  {2\pi \(n(k) - \frac{|\ell(k)|}{2} + \frac{m}{2}\)} + O(\log m)~
\]
where $k={d-c-1\over 2}$.

In summary, we have proved the following.
The condition (\ref{sl2z_bound}) is required for $Z_{NS,+}(\t)$ to be consistent with the phase structure given by the group $\Gamma_\infty\backslash\Gamma_\theta$ (corresponding to distinct Euclidean BTZ black holes which dominate in different regions of parameter space [\onlinecite{Maloney:2007ud}, \S 7.3]).  We have seen that the necessary condition (\ref{deltaonebound}) that we derived earlier in the paper, governing the Hawking-Page transition, is sufficient to guarantee (\ref{sl2z_bound}), and hence the full expected phase diagram.

\section{Estimating the volumes of regions in $\mathbb{RP}^N$}
\label{estimathon}

Recall the problem we have. We would like to estimate
\be
f_N=\frac{\Gamma(\frac{N+1}2)}{\pi^{\frac{N+1}2}}\(\prod_{i=1}^N \int_{-e^{2\pi (n_i - \frac{|\ell_i|}{2}+\frac{m}2)}}^{e^{2\pi (n_i - \frac{|\ell_i|}{2}+\frac{m}2)}} d\xi_i\) \frac{1}{(1+\sum_i(\xi^i)^2)^{\frac{N+1}2}}
\ee
in the large $N$ limit where $N+1$ is the number of polar terms of a Jacobi form of index $m$.

As an example, let's consider $m=2$. We will later switch to the large $m$ limit. There are only two polar terms: $y^2$ and $y^1$ so $N=1$. We normalize the $y^2$ coefficient to $1$, and the coefficient for $y^1$ parametrizes $\mathbb{RP}^1$ (the ``point at infinity" corresponds to a $y^2$ coefficient of $0$, but this has measure zero).

For $y^1$, $\ell=1$ and $n=0$, so
\be
e^{2\pi (n_i - \frac{|\ell_i|}{2}+\frac{m}2)}=e^{\pi}.
\ee
Thus the integral is
\be
f_2 = \frac{\Gamma(1)}{\pi} \int_{-e^{\pi}}^{e^{\pi}} d\xi_1 \frac{1}{(1+\xi_1^2)^1} = 0.9725.
\ee
In the large $m$ limit, there are $\lceil \frac{k+1}2 \rceil$ integrals with limits $-e^{k \pi}$ to $e^{k \pi}$ (see Table II).

\begin{table}[ht]
\caption{Most polar terms at index $m$ (excluding $y^m$).}
\centering
\begin{tabular}{c c}
Term & $e^{2\pi(n-\frac{|\ell|}{2}+\frac{m}{2})}$ \\
\hline
$y^{m-1}$ & $e^{\pi}$ \\
$y^{m-2}$ & $e^{2\pi}$ \\
$q y^m$ & $e^{2\pi}$ \\
$y^{m-3}$ & $e^{3\pi}$ \\
$q y^{m-1}$ & $e^{3\pi}$ \\
$y^{m-4}$ & $e^{4\pi}$ \\
$q y^{m-2}$ & $e^{4\pi}$ \\
$q^2 y^m$ & $e^{4\pi}$ \\
$y^{m-5}$ & $e^{5\pi}$ \\
$q y^{m-3}$ & $e^{5\pi}$ \\
$q^2 y^{m-1}$ & $e^{5\pi}$ \\
\hline
\end{tabular}
\end{table}

\subsection{An Upper Bound}

In this section we will derive an upper bound on $f_N$ of $0.9725$. Recall the famous fact of life that
\be
\frac{\Gamma(\frac{N+2}2)}{\pi^{\frac{N+2}2}}\int_{-\infty}^{\infty} d\xi_{N+1} \frac{1}{(1+\xi_1^2+\ldots+\xi_{N+1}^2)^{\frac{N+2}2}} = \frac{\Gamma(\frac{N+1}2)}{\pi^{\frac{N+1}2}}\frac{1}{(1+\xi_1^2+\ldots+\xi_{N}^2)^{\frac{N+1}2}}.
\label{eq:masterequation}
\ee
Thus, we can always take extra integrals to $\infty$ and we will get a strictly bigger value. No matter how big $N$ is, we will always have an integral with limits $-e^{\pi}$ to $e^{\pi}$ (coming from the $y^{m-1}$ term). In particular
\begin{align}
f_N &= \frac{\Gamma(\frac{N+1}{2})}{\pi^{\frac{N+1}{2}}}\int_{-e^{\pi}}^{e^{\pi}} d\xi_1 \int d\xi_2 \ldots \int d\xi_N \frac{1}{(1+\xi_1^2+\ldots+\xi_N^2)^{\frac{N+1}{2}}} \nn \\
&< \frac{\Gamma(\frac{N+1}{2})}{\pi^{\frac{N+1}{2}}}\int_{-e^{\pi}}^{e^{\pi}} d\xi_1 \int_{-\infty}^{\infty} d\xi_2 \ldots \int_{-\infty}^{\infty} d\xi_N \frac{1}{(1+\xi_1^2+\ldots+\xi_N^2)^{\frac{N+1}{2}}} \nn \\
&= \frac{1}{\pi} \int_{-e^{\pi}}^{e^{\pi}} d\xi_1 \frac{1}{1+\xi_1^2} \nn\\
&= 0.9725.
\end{align}
where we define an integral $d\xi_i$ with unlabelled limits as from $-e^{2\pi (n_i - \frac{|\ell_i|}{2}+\frac{m}2)}$ to $e^{2\pi (n_i - \frac{|\ell_i|}{2}+\frac{m}2)}$.

\subsection{A Lower Bound}

Now we will show a lower bound of $0.9699$ through a series of inequalities. Again we use the same convention of unlabelled limits of integration $d\xi_i$ being from $-e^{2\pi (n_i - \frac{\ell_i^2}{4m}+\frac{m}4)}$ to $e^{2\pi (n_i - \frac{\ell_i^2}{4m}+\frac{m}4)}$.

First we will show
\begin{align}
f_N &=\frac{\Gamma(\frac{N+1}{2})}{\pi^{\frac{N+1}{2}}}\int d\xi_1 \int d\xi_2 \ldots \int d\xi_N \frac{1}{(1+\xi_1^2+\ldots+\xi_N^2)^{\frac{N+1}{2}}} \nn \\
&> 1 - \sum_{i=1}^N \(1 - \frac{1}{\pi} \int d\xi_i \frac{1}{1+\xi_i^2}\).
\label{eq:goal}
\end{align}
To see this, first rewrite (\ref{eq:goal}) as
\begin{align}
1- \frac{\Gamma(\frac{N+1}{2})}{\pi^{\frac{N+1}{2}}}\int d\xi_1 \int d\xi_2 \ldots \int d\xi_N \frac{1}{(1+\xi_1^2+\ldots+\xi_N^2)^{\frac{N+1}{2}}} \nn \\
< \sum_{i=1}^N \(1 - \frac{1}{\pi} \int d\xi_i \frac{1}{1+\xi_i^2}\).
\label{eq:rewrite}
\end{align}
Now note that
\be
\frac{1}{\pi} \int d\xi_i \frac{1}{1+\xi_i^2}
\ee
represents the fraction of $\mathbb{RP}^N$ where $\xi_i$ is between the appropriate limits of $-e^{2\pi (n_i - \frac{|\ell_i|}{2}+\frac{m}2)}$ to $e^{2\pi (n_i - \frac{|\ell_i|}{2}+\frac{m}2)}$. We write this as a fraction of $\mathbb{RP}^N$ instead of $\mathbb{RP}^1$ by using (\ref{eq:masterequation}) to add the remaining $N-1$ integrals from $-\infty$ to $\infty$ and change the prefactor.

In more detail, let's take the first term ($i=1$) in the sum in the right hand side of (\ref{eq:rewrite}). That term, in the large $N$ limit, is
\begin{align}
1-\frac{1}{\pi} \int_{-e^{\pi}}^{e^{\pi}} d\xi_1 \frac{1}{1+\xi_1^2} &= 1-\frac{\Gamma(\frac{N+1}2)}{\pi^{\frac{N+1}2}} \int_{-e^{\pi}}^{e^{\pi}} d\xi_1 \int_{-\infty}^{\infty} d\xi_2 \ldots \int_{-\infty}^{\infty} d\xi_N \frac{1}{(1+\xi_1^2+\ldots+\xi_N^2)^{\frac{N+1}{2}}}
\label{eq:new}
\end{align}
which is exactly the region \emph{outside} $-e^{\pi} < \xi_1 < e^{\pi}$ in $\mathbb{RP^N}$. However, the left hand side of (\ref{eq:rewrite}) is $\mathbb{RP^N}$ with the region
$$
|\xi_i| \leq R(i), \phantom{a} \forall \phantom{a} i
$$
excluded.

Thus, (\ref{eq:rewrite}) is satisfied by using the fact that the complement of the intersection is less than the sum of complements.
We illustrate this for $N=2$ in Figure 9 using a diagram by the famous Dr. John Venn.

\begin{figure}
\begin{center}
\includegraphics[width=0.40\textwidth]{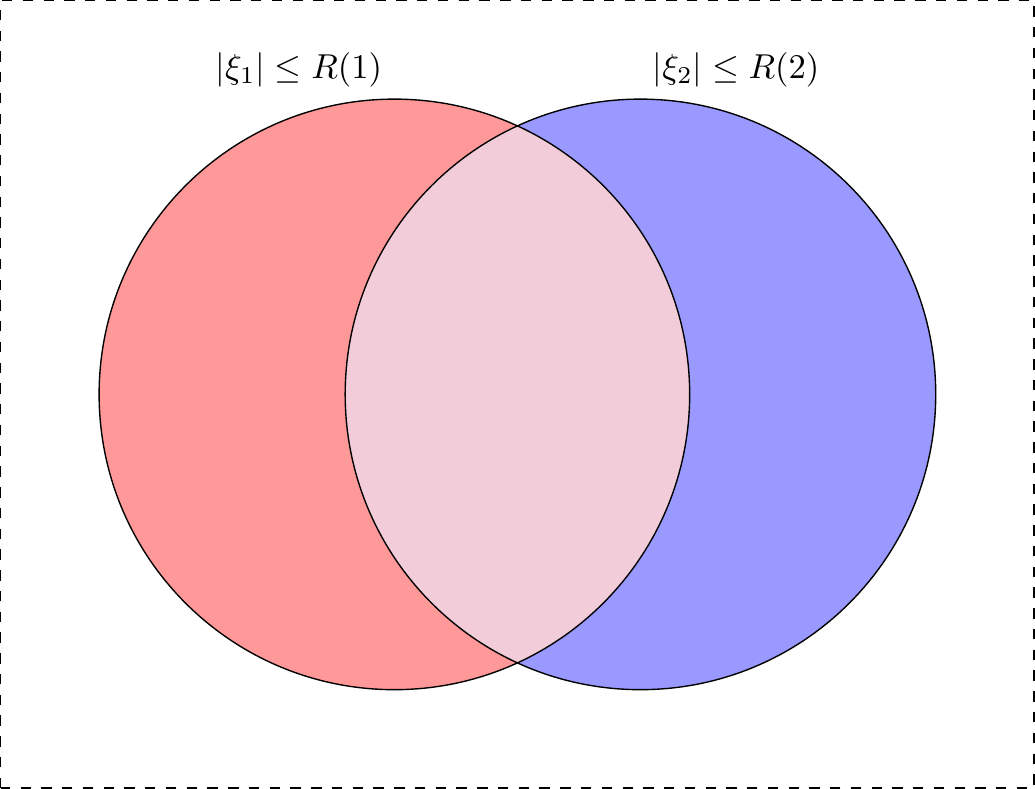}
\end{center}
\caption{
Here, we show an image from Dr. John Venn [\onlinecite{venn}]. The LHS of (\ref{eq:rewrite}) for $N=2$ is represented by the white, blue, and red parts. The RHS is the sum of two terms: blue and white, and red and white. Thus, the sum is greater the the LHS.}
\end{figure}

Now we are in business. It is another classic fact of life that
\begin{align}
\frac{1}{\pi} \int_{-R(i)}^{R(i)} d\xi_i \frac{1}{1+\xi_i^2} &= \frac{2}{\pi} \arctan{R(i)} \nn \\
&= 1 - \frac{2}{\pi R(i)} + \ldots
\end{align}
Plug into (\ref{eq:goal}), to get:
\begin{align}
1 - \sum_{i=1}^N \(1 - \frac{1}{\pi} \int d\xi_i \frac{1}{1+\xi_i^2}\) > 1 - \sum_{i=1}^{N} \frac{2}{\pi R(i)}
\end{align}
In the large $m$ limit, the first terms look like
\begin{align}
1 - \sum_{i=1}^{N} \frac{2}{\pi a_i} &= 1 - \frac{2}{\pi} \( \frac{1}{e^\pi} + \frac{2}{e^{2\pi}} + \frac{2}{e^{3\pi}} + \frac{3}{e^{4\pi}} + \frac{3}{e^{5\pi}} + \ldots \) \nn\\
&> 1 - \frac{2}{\pi} \( \frac{1}{e^{\pi}} + \frac{2}{e^{2\pi}} + \frac{3}{e^{3\pi}} + \frac{4}{e^{4\pi}} + \frac{5}{e^{5\pi}} + \ldots \) \nn \\
&=1 - \frac{2}{\pi} \( \frac{1}{e^{\pi}(1-\frac{1}{e^{\pi}})^2}\) \nn \\
&= 0.9699.
\end{align}

Thus, putting everything together, we get
\be
f_N > 0.9699.
\ee

\end{document}